\documentclass[manuscript]{aastex}
\shorttitle{13 UGC galaxies with Chandra and VLA}
\shortauthors{Kharb et al.}
\begin{document}
\title{Examining the Radio-Loud/Radio-Quiet dichotomy with new Chandra and VLA observations of 13 UGC galaxies}
\author{P. Kharb\altaffilmark{1}} 
\affil{$^{1}$Physics Department, Rochester Institute of Technology, Rochester, NY 14623} 
\email{kharb@cis.rit.edu} 
\author{A. Capetti\altaffilmark{2}}
\affil{$^{2}$INAF - Osservatorio Astronomico di Torino, Strada Osservatorio 20, 10025 Pino Torinese, Italy}
\author{D. J. Axon\altaffilmark{3,1}}
\affil{$^{3}$School of Mathematical and Physical Sciences, Univ. of Sussex, Brighton BN1 9QH, UK}
\author{M. Chiaberge\altaffilmark{4, 6, 9}}
\affil{$^{4}$Space Telescope Science Institute, 3700 San Martin Drive, Baltimore, MD 21218}
\affil{$^{9}$Center for Astrophysical Sciences, Johns Hopkins University, 3400 N. Charles Street Baltimore, MD 21218}
\author{P. Grandi\altaffilmark{5}}
\affil{$^{5}$INAF - Istituto di Astrofisica Spaziale e Fisica Cosmica, Bologna, Italy}
\author{A. Robinson\altaffilmark{1}}
\author{G. Giovannini\altaffilmark{6}}
\affil{$^{6}$INAF - Istituto di Radioastronomia di Bologna, via Gobetti 101, 40129 Bologna, Italy}
\author{B. Balmaverde\altaffilmark{7}}
\affil{$^{7}$Universita di Torino, Via P. Giuria 1, Torino I-10125, Italy}
\author{D. Macchetto\altaffilmark{4}}
\author{R. Montez\altaffilmark{8}}
\affil{$^{8}$Center for Imaging Science, Rochester Institute of Technology, Rochester, NY 14623} 

\begin{abstract}
We present the results from new $\sim$15 ks {\it Chandra}-ACIS and 4.9 GHz Very Large Array observations of 13 galaxies hosting low luminosity AGN. This completes the multiwavelength study of a sample of 51 nearby early-type galaxies described in \citet{Capetti05,CapettiBalmaverde06,BalmaverdeCapetti06}. The aim of the three previous papers was to explore the connection between the host galaxies and AGN activity in a radio-selected sample. We detect nuclear X-ray emission in eight sources and radio emission in all but one ({\it viz.}, UGC\,6985). The new VLA observations improve the spatial resolution by a factor of ten: the presence of nuclear radio sources in 12 of the 13 galaxies confirms their AGN nature. As previously indicated, the behavior of the X-ray and radio emission in these sources depends strongly on the form of their optical surface brightness profiles derived from {\it Hubble Space Telescope} imaging, {\it i.e.,} on their classification as ``core'', ``power-law'' or ``intermediate'' galaxies. With more than twice the number of ``power-law'' and ``intermediate'' galaxies compared to previous work, we confirm with a much higher statistical significance that these galaxies lie well above the radio-X-ray correlation established in FRI radio galaxies and the low-luminosity ``core'' galaxies. This result highlights the fact that the ``radio-loud/radio-quiet'' dichotomy is a function of the host galaxy's optical surface brightness profile. We present radio-optical-X-ray spectral indices for all 51 sample galaxies. Survival statistics point to significant differences in the radio-to-optical and radio-to-X-ray spectral indices between the ``core'' and ``power-law'' galaxies (GehanÕs Generalized Wilcoxon test probability $p$ for the two classes being statistically similar is $<10^{-5}$), but not in the optical-to-X-ray spectral indices ($p= 0.25$). Therefore, the primary difference between the ``core'' and ``power-law'' galaxies is in their ability to launch powerful radio outflows. This result is consistent with the {hypothesis} of different formation processes and evolution histories in ``core'' and ``power-law'' galaxies: major mergers are likely to have created ``core'' galaxies, while minor mergers were instrumental in the creation of ``power-law'' galaxies. 
\end{abstract}

\section{Introduction}
It is now believed that most, if not all, galaxies host a supermassive black hole (SMBH, $10^{7-9}M_{\sun}$) in their centers \citep{KormendyRichstone95}. The tight relationship between the SMBH mass and the stellar velocity dispersion \citep[e.g.,][]{FerrareseMerritt00,Gebhardt00} indicates that they follow a common evolutionary path. \citet{Heckman04} have demonstrated that the synchronous growth of SMBH and galaxies is in progress in the local Universe. However, a clear picture of the relationship between an active galactic nucleus (AGN) and its host galaxy is yet to emerge. On the basis of the ratio ($R$) between the radio (1.4 GHz) and optical ($B$-band) flux density, AGNs can be broadly divided into the radio-loud ($R\ge10$) {and} radio-quiet ($R<$10) categories \citep{Kellermann89}. While nearby radio-quiet AGNs appear to typically reside in spiral galaxies, elliptical galaxies can host both radio-loud and radio-quiet AGNs \citep[e.g.,][]{Ho02,Dunlop03}. Furthermore, while there is a median shift between the SMBH mass distributions of radio-quiet and radio-loud AGNs, with the radio-loud AGN being generally associated with the most massive SMBH, both distributions are broad and overlap considerably.

In recent years, a new picture of the host galaxies has emerged through high angular resolution observations with the {\it Hubble Space Telescope} (HST). Nearly all galaxies have singular starlight distributions with surface brightness profiles (SBP) diverging as $\Sigma(r)\sim r^{-\gamma}$, with $\gamma>0$ \citep[{the ``Nuker law''},][]{Lauer95}, and the distribution of cusp slopes is bimodal \citep[e.g.,][]{Gebhardt96, Faber97}. In some galaxies, the projected profile breaks into a shallow inner cusp with $\gamma<0.3$ and these form the class of ``core'' galaxies. In other galaxies $\gamma>0.5$ down to the HST resolution limit; these systems are classified as ``power-law'' galaxies. Only a small number of ``intermediate'' galaxies have been identified with $0.3<\gamma<0.5$ \citep{Ravindranath01}. The central structure correlates with other galaxy properties, showing that luminous early-type galaxies preferentially have ``cores", whereas most fainter spheroids have ``power-law'' profiles. Moreover, ``core'' galaxies rotate slowly and have boxy isophotes, while ``power-law'' galaxies rotate rapidly and are disky \citep{Faber97}.

With the aim of examining the pertinent issue of the AGN-host galaxy connection, a systematic study of nearby early-type galaxies was initiated by \citet{Capetti05,BalmaverdeCapetti06,CapettiBalmaverde06} (Papers I, II and III, respectively). The initial sample comprised of 65 galaxies with {HST} archival data that had radio detections in the CfA redshift survey \citep[48 sources,][]{wrobel91a} and the Parkes survey \citep[17 sources,][]{Sadler89}. However, only 51 of these 65 galaxies could be classified as ``core'', ``power-law" or ``intermediate'' on the basis of the ``Nuker law'' surface brightness fit (see Paper I). 

\citet{graham03} have argued that a S\'ersic model \citep{sersic65,sersic68} provides a better characterization of the brightness profiles of early-type galaxies. They have also suggested a new definition of a ``core'' galaxy as the class of objects showing a light deficit toward the center with respect to the S\'ersic law \citep{trujillo04}. 
{We note that for nearby luminous bright galaxies, the two surface brightness classification schemes however, provide a consistent classification, {\it i.e.,} the objects classified as ``core'' galaxies in the Nuker scheme \citep[e.g.,][]{Lauer95}, are recovered as such with the Graham et al. definition (see Papers I, II).}

Using the Very Large Array (VLA) and {\it Chandra} archival data, a radio-X-ray study of 38 galaxies was carried out in Papers II and III. This study found that the ``power-law'' and ``core'' galaxies covered different regions of the X-ray-Radio luminosity plane. The ``core'' galaxies extended the well-known correlation between nuclear radio and X-ray luminosity in the Fanaroff-Riley type I (FRI) radio galaxies, supporting the interpretation that the ``core'' galaxies represented the scaled down, low luminosity analogues of radio-loud AGN \citep[e.g.,][]{CapettiKharb09}. The FRI correlation itself is indicative of a common non-thermal AGN jet origin of the radio, X-ray and optical nuclear emission \citep{Chiaberge99,Hardcastle00,Kharb04}. The ``power-law" galaxies showed a deficit in radio luminosity at a given X-ray luminosity (or an excess in the X-ray emission at a given radio luminosity) with respect to the ``core'' galaxies by a factor of $\sim 100$. The (two) ``intermediate'' galaxies lay in the ``power-law'' region. The ``power-law" and ``intermediate'' galaxies showed nuclear radio and X-ray luminosities similar to those measured in Seyfert and Low-ionization nuclear emission-line region (LINER) galaxies, thereby suggesting that they were associated with genuine low luminosity active nuclei, representing the local manifestation of radio-quiet AGN.

However, a potential weakness of the previous analysis was that the sample was biased towards ``core'' galaxies ($\sim75$\% of the sample), driven primarily by the availability of {\it Chandra} archival X-ray data for the sample sources. There were only nine ``power-law/intermediate'' galaxies. In order to put these results on a firmer statistical footing, we obtained new {\it Chandra} X-ray observations for the 13 Uppsala General Catalogue (UGC) galaxies which did not have {\it Chandra} archival data. Furthermore, we obtained VLA {A-array} 4.9 GHz radio data, improving the spatial resolution from 5\arcsec\ of the original Wrobel's survey to 0\farcs5, increasing substantially our ability to isolate the genuine nuclear radio emission. 

{We present here the results of these new {\it Chandra} and VLA observations.} With these new data, we have now {\it more than doubled} the number of ``power-law'' and ``intermediate'' galaxies observed. The paper is arranged as follows: \S2 and \S3 describe the X-ray and radio observations and data-analysis, respectively. The results, discussion and summary follow in \S4, \S5 and \S6. The spectral index $\alpha$ is defined such that flux density at frequency $\nu$ is, $S_{\nu}\propto\nu^{-\alpha}$ and the photon index is $\Gamma=1+\alpha$. To be consistent with Papers I $-$ III, we have adopted a cosmology in which $H_0$ = 75 km~s$^{-1}$~Mpc$^{-1}$, $q_{0}=0.5$. {At the distance of the nearest (UGC\,6985, $z$=0.0031) and the farthest (UGC\,10656, $z$=0.0093) sample galaxy discussed here, 1$\arcsec$ corresponds to 72 and 178 parsec, respectively.}

\section{Radio Observations and Data Analysis}
The radio observations were carried out with the VLA at 4.86 GHz in the A-array configuration on 2006, February 24 (Project code: AC807). The effective spatial resolution of the radio observations was $\sim0\farcs5$, which typically translates to  scales of $\sim$50 pc at the distance of the sources. The data were acquired using two intermediate frequency (IF) channels in dual polarization mode with a bandwidth of 50 MHz. 3C\,286 was used as primary flux calibrator. The sources were observed in single scans of duration 15$-$20 mins, adjacent in time to $\sim$2.5 mins scans of a nearby bright phase-calibrator. The data reduction was carried out following standard calibration and reduction procedures in the Astronomical Image Processing System (AIPS). After the initial amplitude and phase calibration, the AIPS tasks CALIB and IMAGR were used iteratively to self-calibrate and image the sources. Radio emission was detected from the centers of all but one galaxy ({\it viz.,} UGC\,6985). The radio images are presented in Figure \ref{figONE}. {The final {\it rms} noise in the radio images is typically of the order of $\sim0.05$ mJy\,beam$^{-1}$.}

The AIPS tasks TVWIN and IMSTAT were used to derive the integrated radio flux densities (see Table~1). Three sources {\it viz.,} UGC\,0968, UGC\,5959 and UGC\,12759, were not observed as part of the program AC807. The 4.9 GHz flux density for UGC\,5959 was obtained from \citet{wrobel91a}, while the estimates for UGC\,0968 and UGC\,12759 were obtained from Henrique Schmitt (private communication).

With respect to the data presented by \citet{wrobel91b,wrobel91a}, the current VLA data has improved the spatial resolution from $\sim$ 5\arcsec\ to $\sim$ 0\farcs5. Nonetheless, the derived radio flux densities are generally in good agreement with most sources differing by less than a factor of 2 (see Table \ref{tabsample}). In several instances (but especially in UGC\,6946), the A array flux densities are higher, possibly indicative of variability and the AGN nature of the emitting sources. The most notable exception is the sole undetected source ({\it viz.}, UGC\,6985) in our data, where the 3$\sigma$ upper limit of 0.12 mJy is $>$10 times fainter than {the flux density of 1.4 mJy reported in the literature.} This indicates that most of the flux measured at the lower resolution is resolved out and it is not associated with a nuclear point source. Multifrequency VLA radio observations would prove valuable {in obtaining the} radio spectral indices and truly {confirming} the AGN nature of these radio nuclei.

\section{X-ray Observations and Data Analysis}
The 13 UGC galaxies were observed with the {\it Chandra} X-ray Observatory from November 2005 to November 2006 for $\sim$15 ks each. The observations were obtained using the AXAF CCD Imaging Spectrometer (ACIS) S3 chip (back illuminated for low-energy response) in the very faint (VFAINT) timed mode. In order to reduce the effect of pileup of the galaxy cores, the observations were carried out using the 1/8 subarray mode (frame time = 0.441 sec). The data were reduced and analysed using the CIAO\footnote{http://cxc.harvard.edu/ciao/} software version 4.1 and calibration database (CALDB) version 4.1.2. Using the CIAO tool ``$acis\_process\_events$'' we reprocessed the level 1 data to remove the 0.5 pixel randomization and do the VFAINT mode background cleaning for all the sources except UGC\,6946 which appeared to be affected by pileup in the nuclear region \citep{Davis01}. The data were filtered on the ASCA grades 0, 2, 3, 4, 6 and on status. Finally, the Good Time Intervals (GTI) supplied with the pipeline were applied to create the new level 2 events file.

While X-ray emission is detected from the central regions of all 13 UGC galaxies, compact nuclear emission is detected in only eight of them. Figure \ref{figONE} displays the
energy filtered 0.5$-$7.0 keV X-ray images, which were binned to 0.25 of the native pixel size and smoothed with a Gaussian of kernel radius = 3, in the astronomical imaging and data visualization application, DS9. The CIAO tool ``{\it specextract}'' was used to extract the spectrum from the nuclear regions. The nuclear region was taken to be a circular region with radius equal to 2$\arcsec$, centered on the pixel with the highest number of counts. The background was chosen as an annular region surrounding the nucleus with radii between 2$\arcsec-4\arcsec$. For sources with greater than 100 counts, a {spectral} binning of 25 counts~bin$^{-1}$ was used, while for sources with fewer than 100 counts, a {spectral} binning of 10 counts~bin$^{-1}$ was used. For the bright source, UGC\,6946, a {spectral} binning of 80 counts~bin$^{-1}$ was used. The spectral fitting and analysis was done using the XSPEC software ({\it HEASOFT version 6.6.3}), using events with energy $>$0.5 keV, where the calibration is well known.

We modelled the grouped spectra with an absorbed powerlaw model (XSPEC model ={\it phabs}({\it powerlaw})). However, for all but one source ({\it viz.,} UGC\,6946), the model-fits were not well constrained. Therefore, for these sources, which typically had fewer than 100 - 150 counts, we obtained unabsorbed fluxes for the power-law model by keeping the neutral hydrogen column density fixed to the Galactic value {toward} the source, and the photon index fixed to 1.7. Fluxes were also obtained for photon indices 1.5 and 2.0, to get error estimates for the flux densities related to the uncertainty on $\Gamma$. 

As the bright source UGC\,6946 was significantly affected ($\approx25\%$) by pileup in its nucleus, we obtained the pileup fraction by following the relevant thread in {\it Sherpa}\footnote{See http://cxc.harvard.edu/sherpa/threads/pileup/}. Subsequently, the pileup-corrected nuclear flux for an absorbed powerlaw model was obtained in XSPEC (model = {\it pileup}({\it phabs}({\it powerlaw})) $-$ flux was obtained after removing the pileup component). 

\begin{figure}
\includegraphics[width=9.2cm]{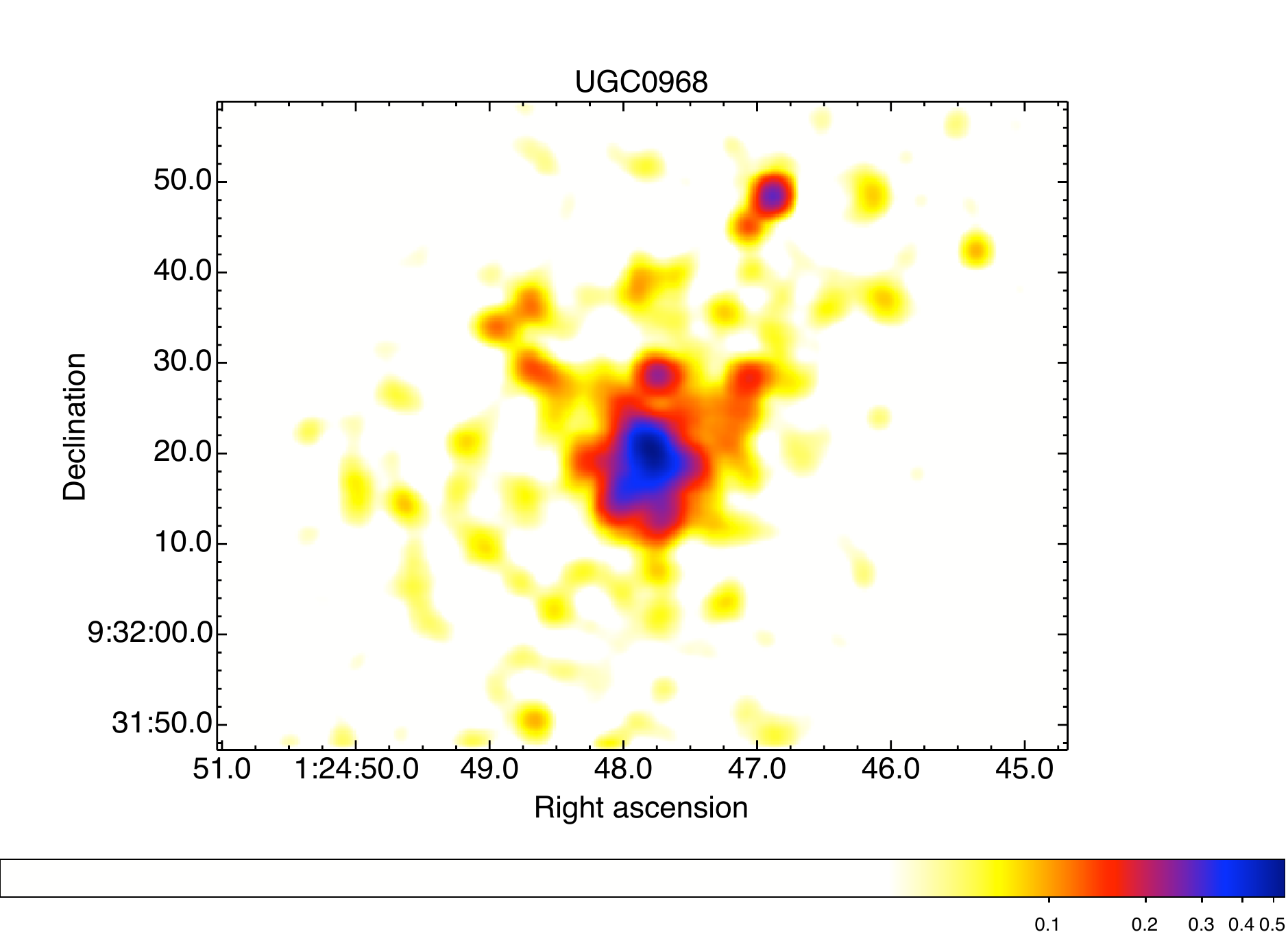}
\includegraphics[width=9.2cm]{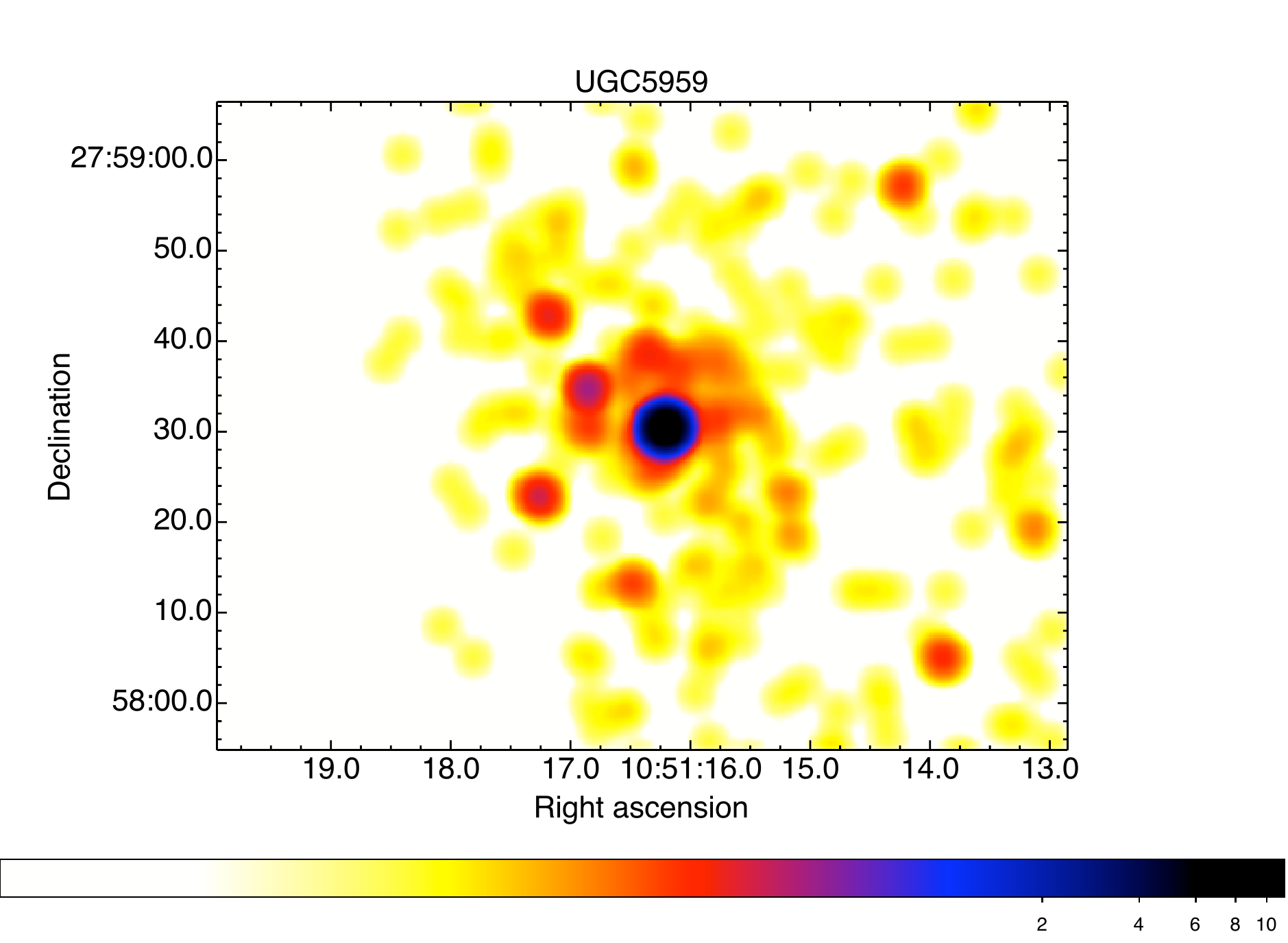}
\includegraphics[width=9.2cm]{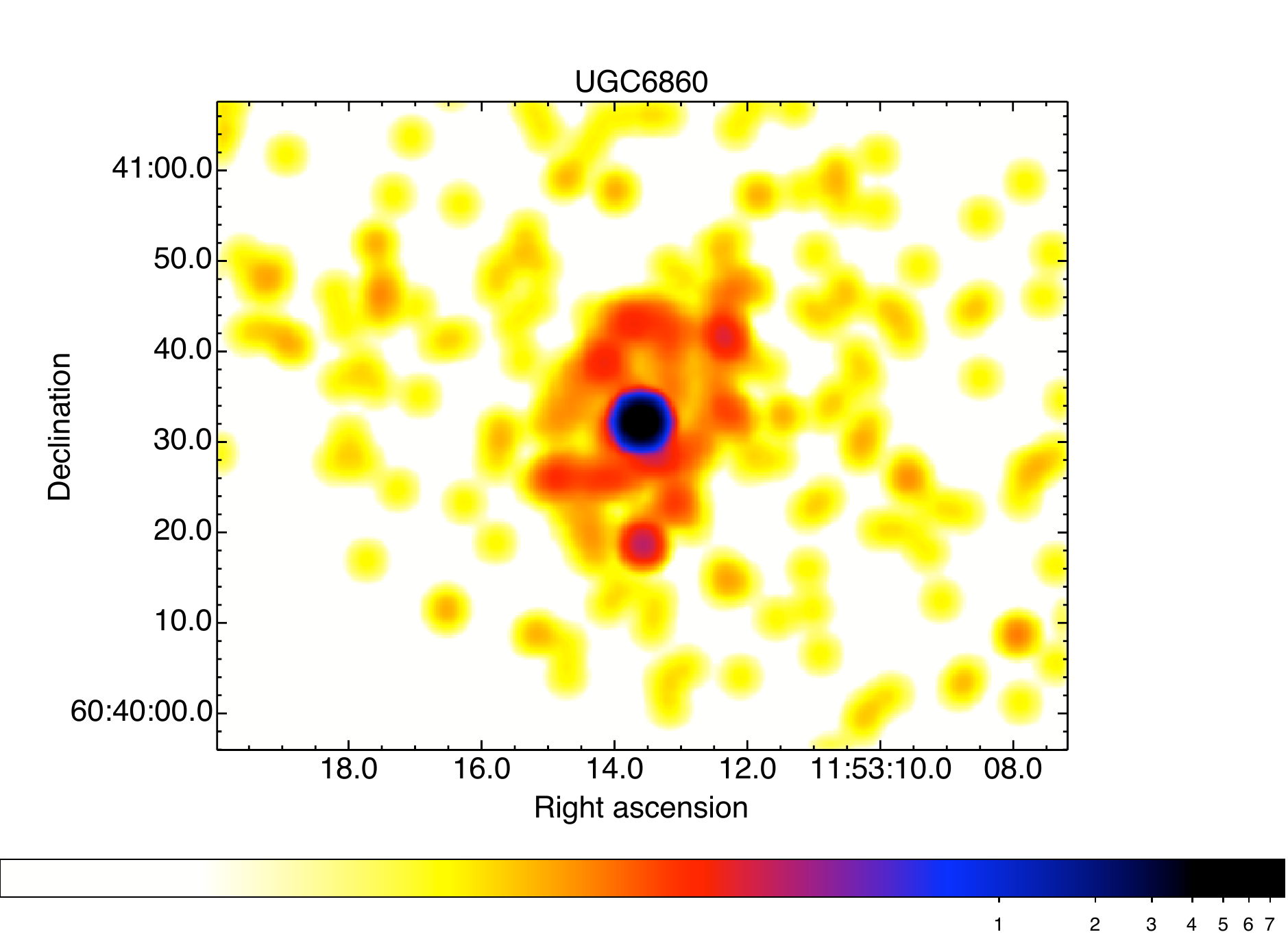}
\includegraphics[width=7.6cm]{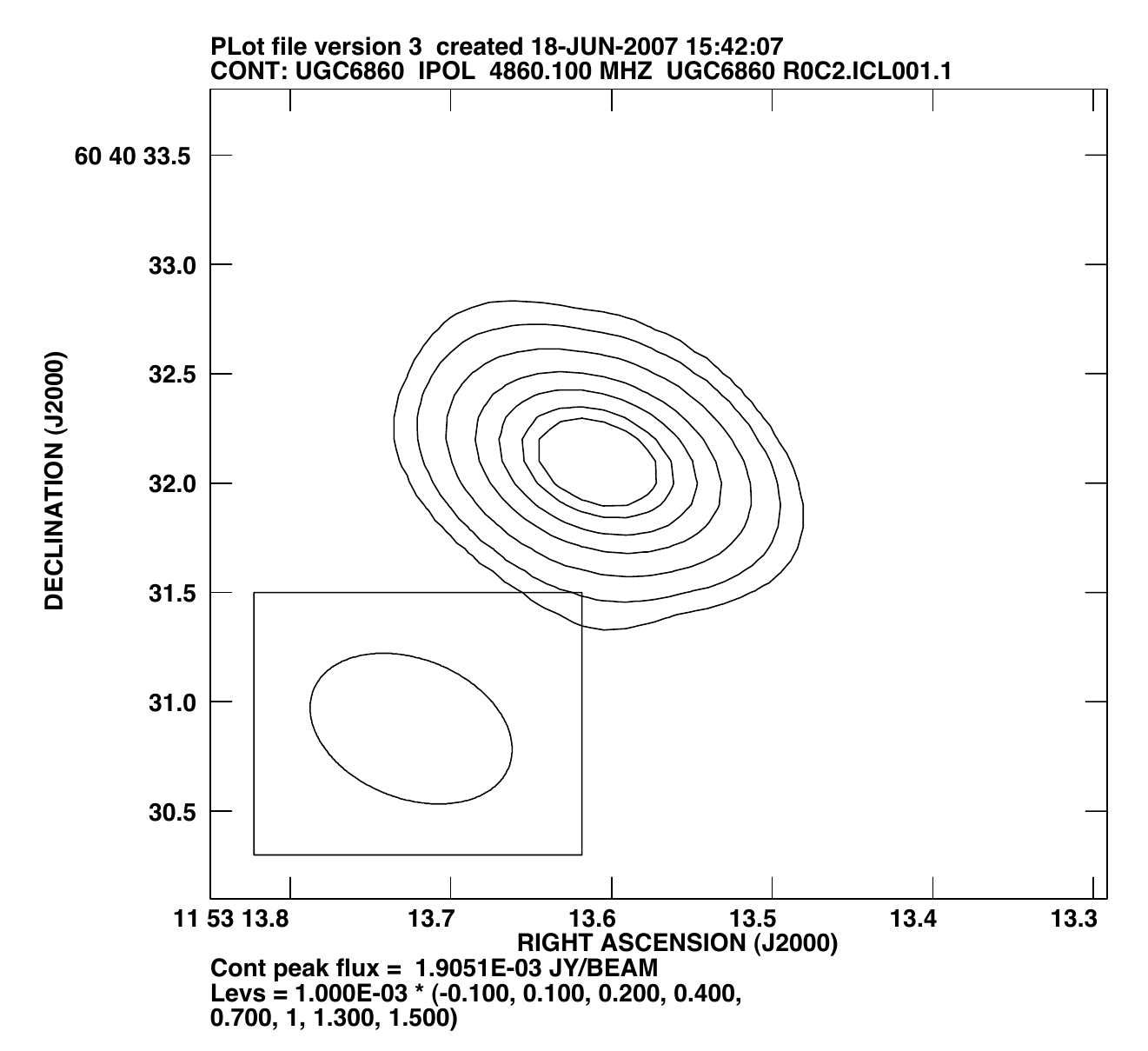}
\caption{\small Energy filtered (0.5$-$7.0 keV) {\it Chandra} color images, and VLA 4.9 GHz radio contour maps. The X-ray images were binned to 0.25 of the native pixel size and smoothed with a Gaussian of kernel radius = 3. The color scale displays the X-ray intensity range in counts~arcsec$^{-2}$. The radio contour levels {listed at the bottom of the radio panels} are in units of Jy~beam$^{-1}$.} 
\label{figONE}
\end{figure}
\addtocounter{figure}{-1}
\begin{figure}
\includegraphics[width=9.2cm]{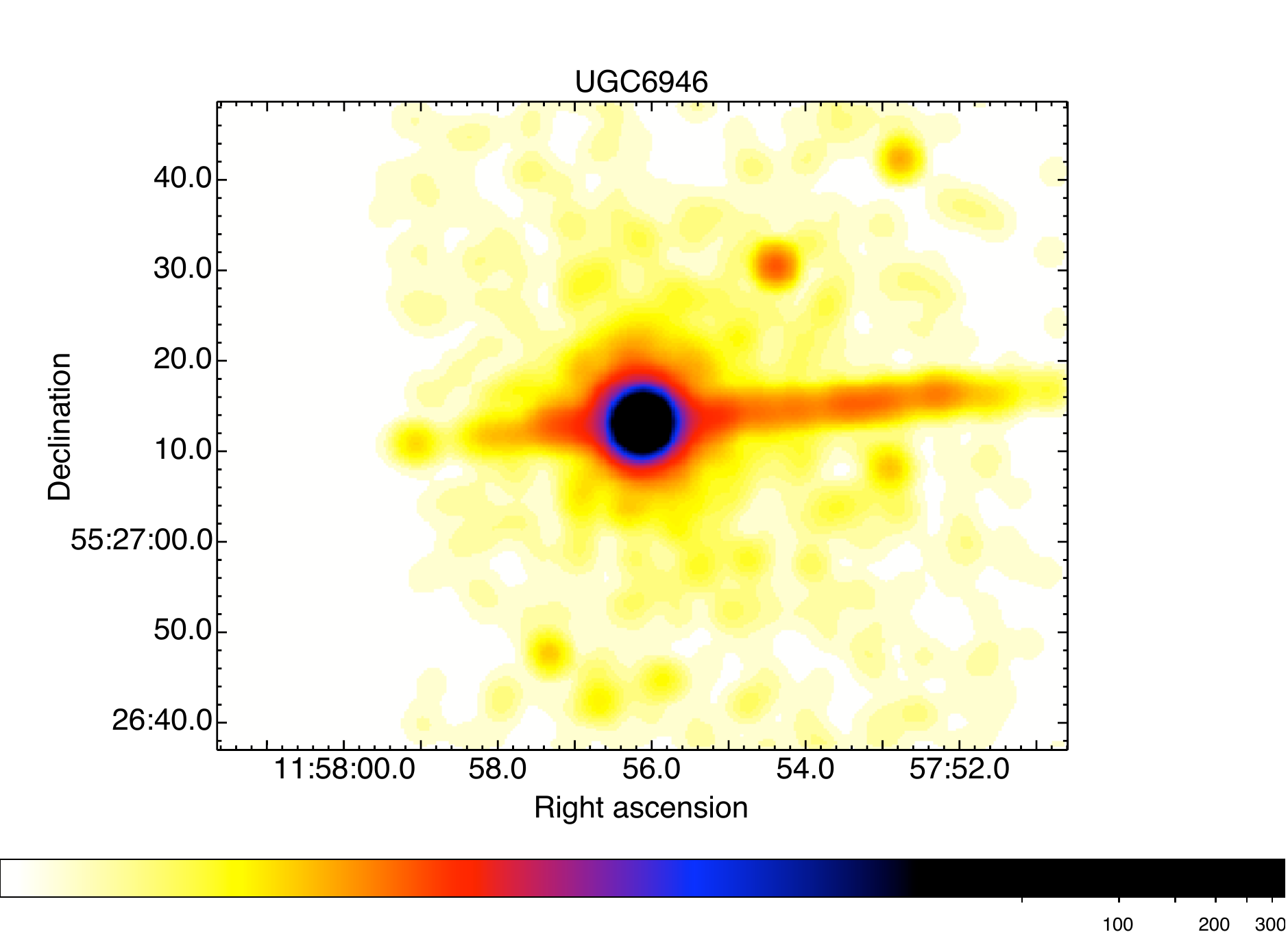}
\includegraphics[width=7.6cm]{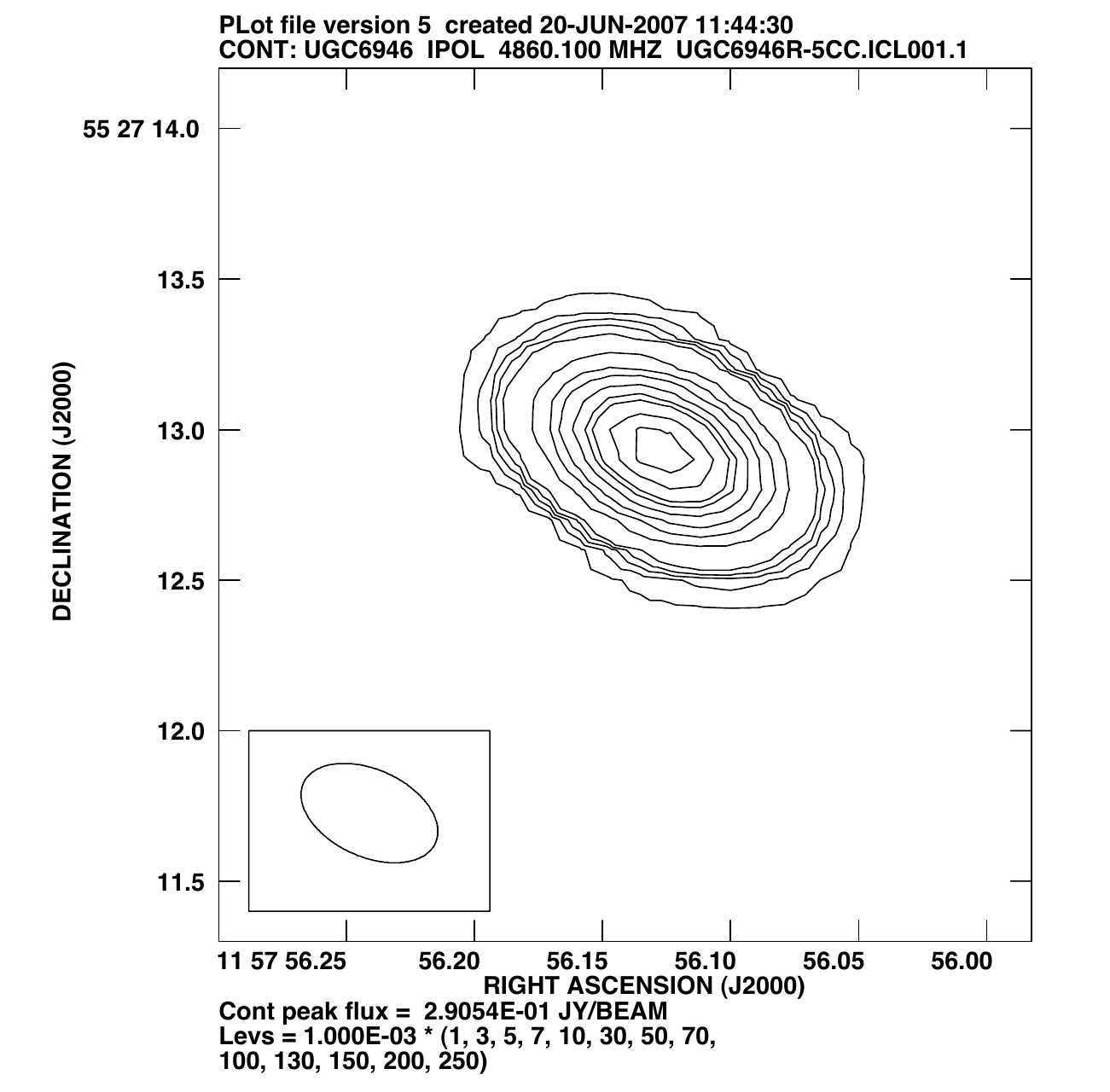}
\includegraphics[width=9.2cm]{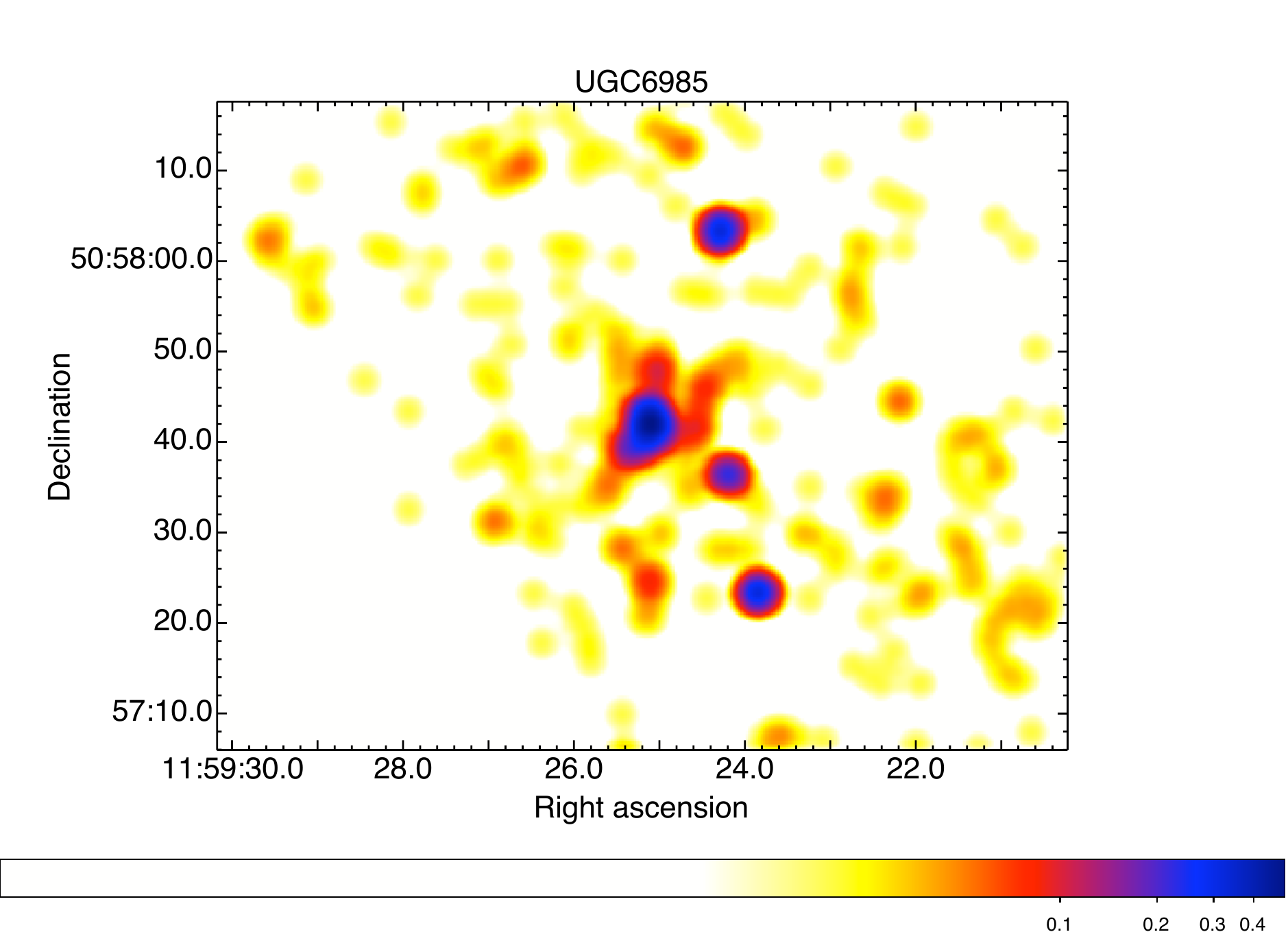}
\includegraphics[width=9.2cm]{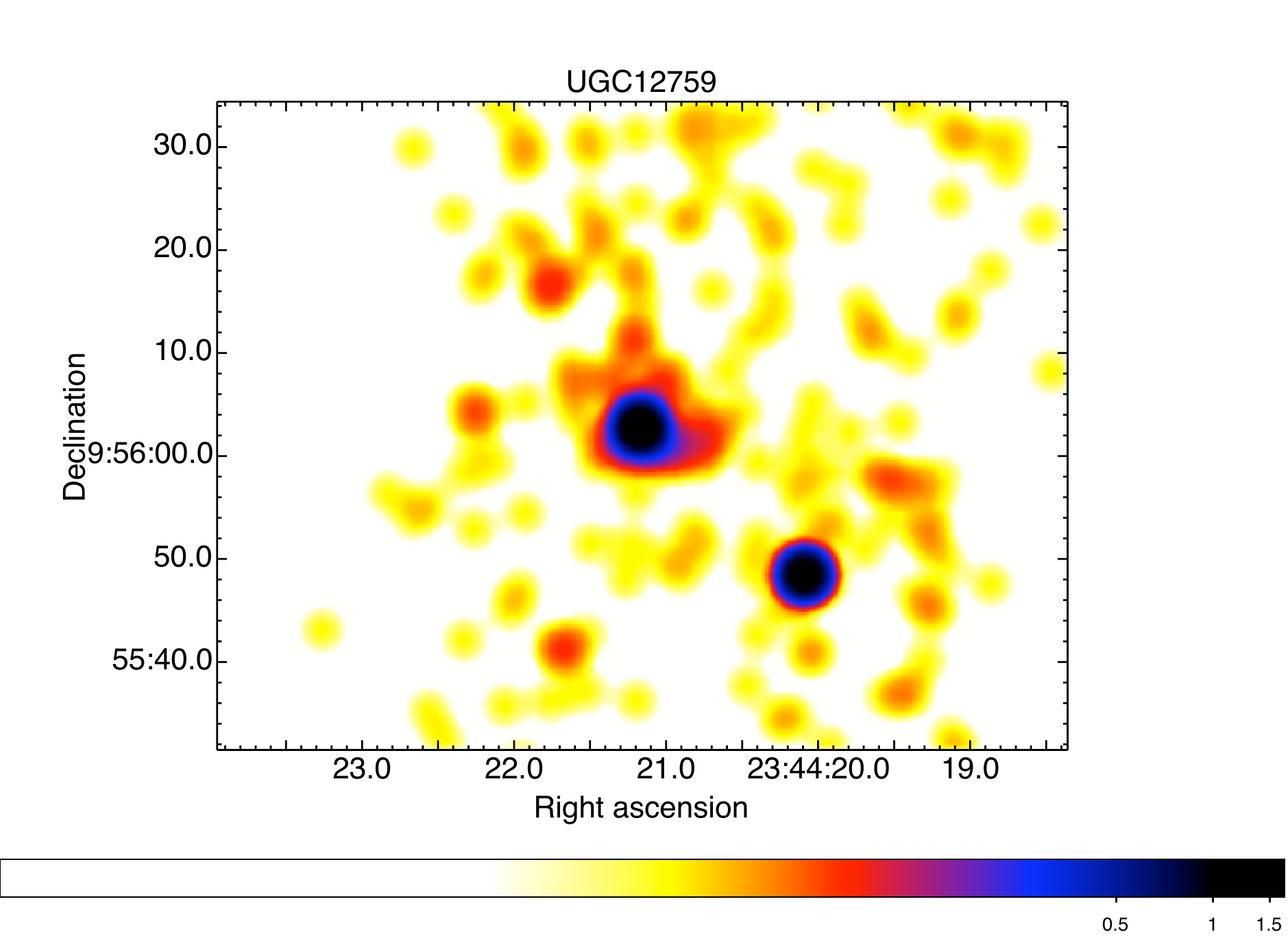}
\caption{\small (continued) }
\end{figure}
\addtocounter{figure}{-1}
\begin{figure}
\includegraphics[width=9.2cm]{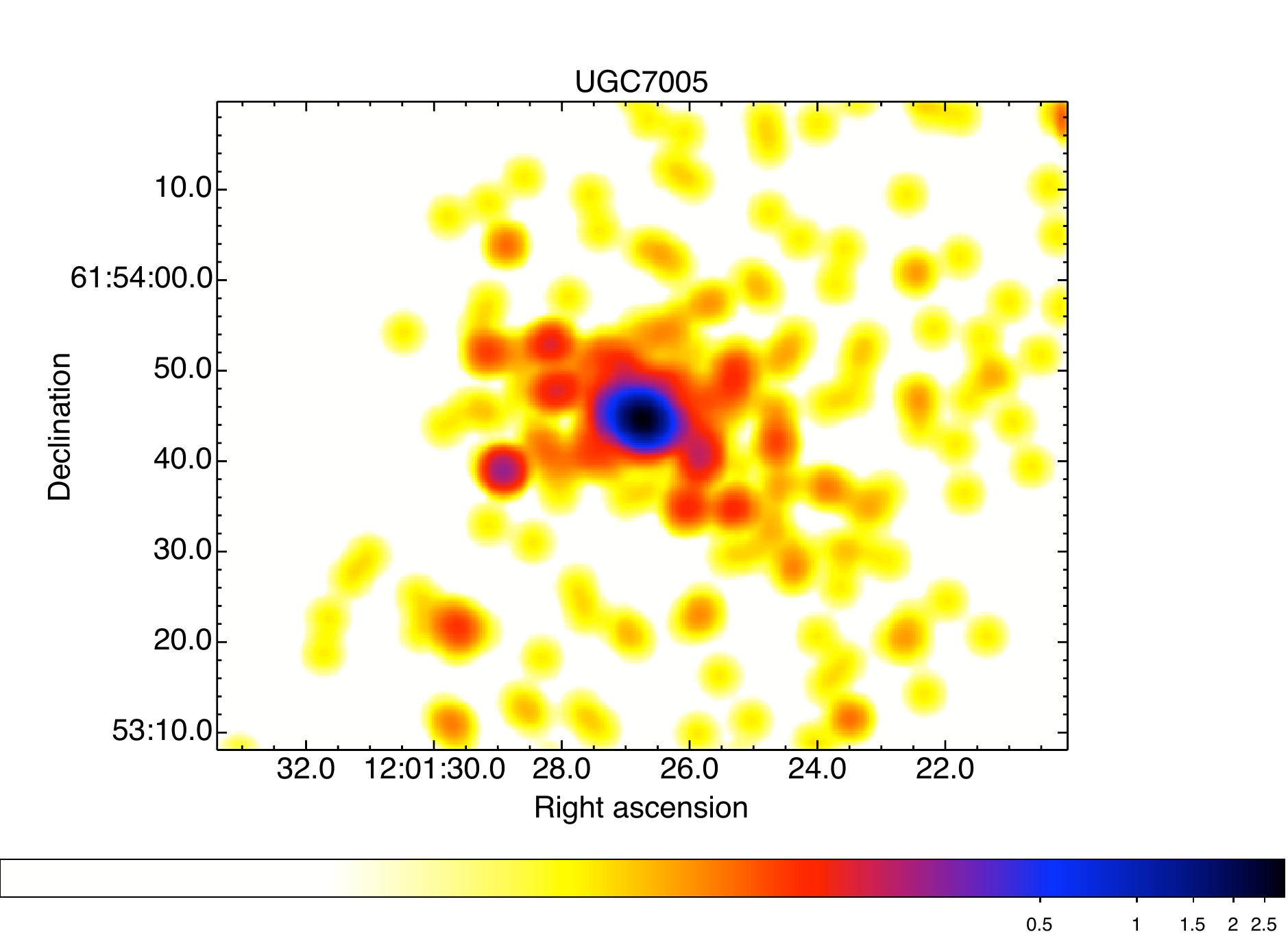}
\includegraphics[width=7.6cm]{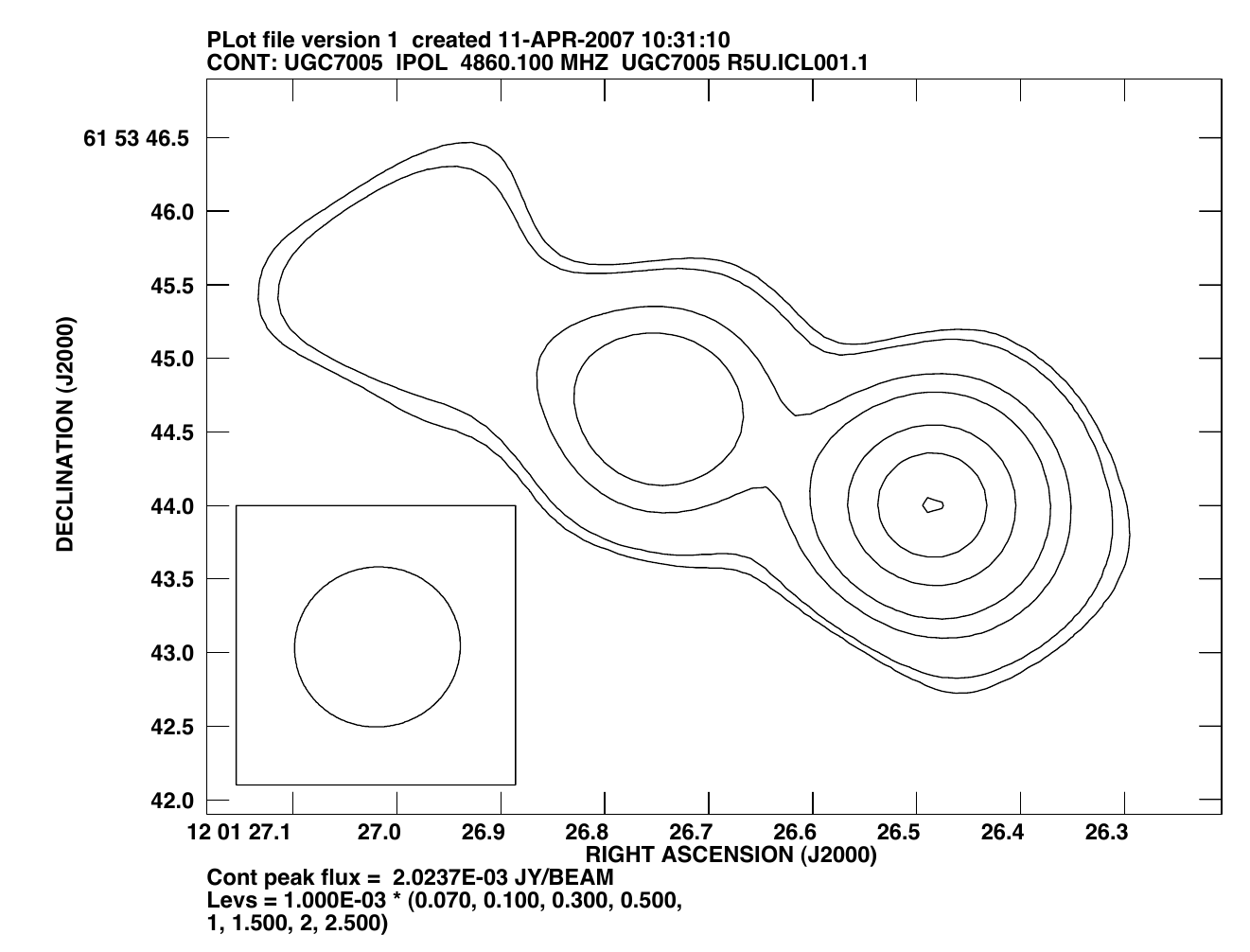}
\includegraphics[width=9.2cm]{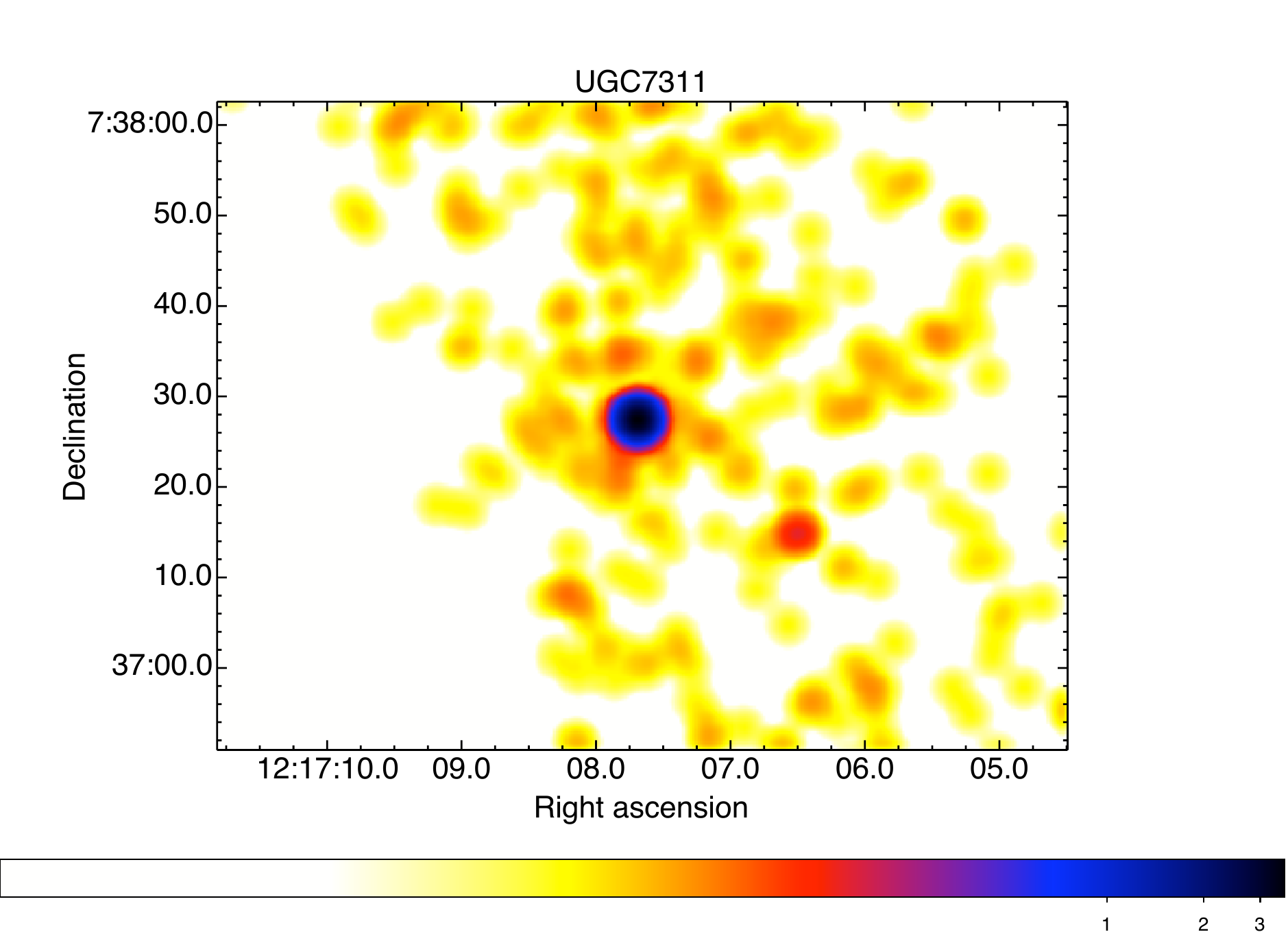}
\includegraphics[width=7.6cm]{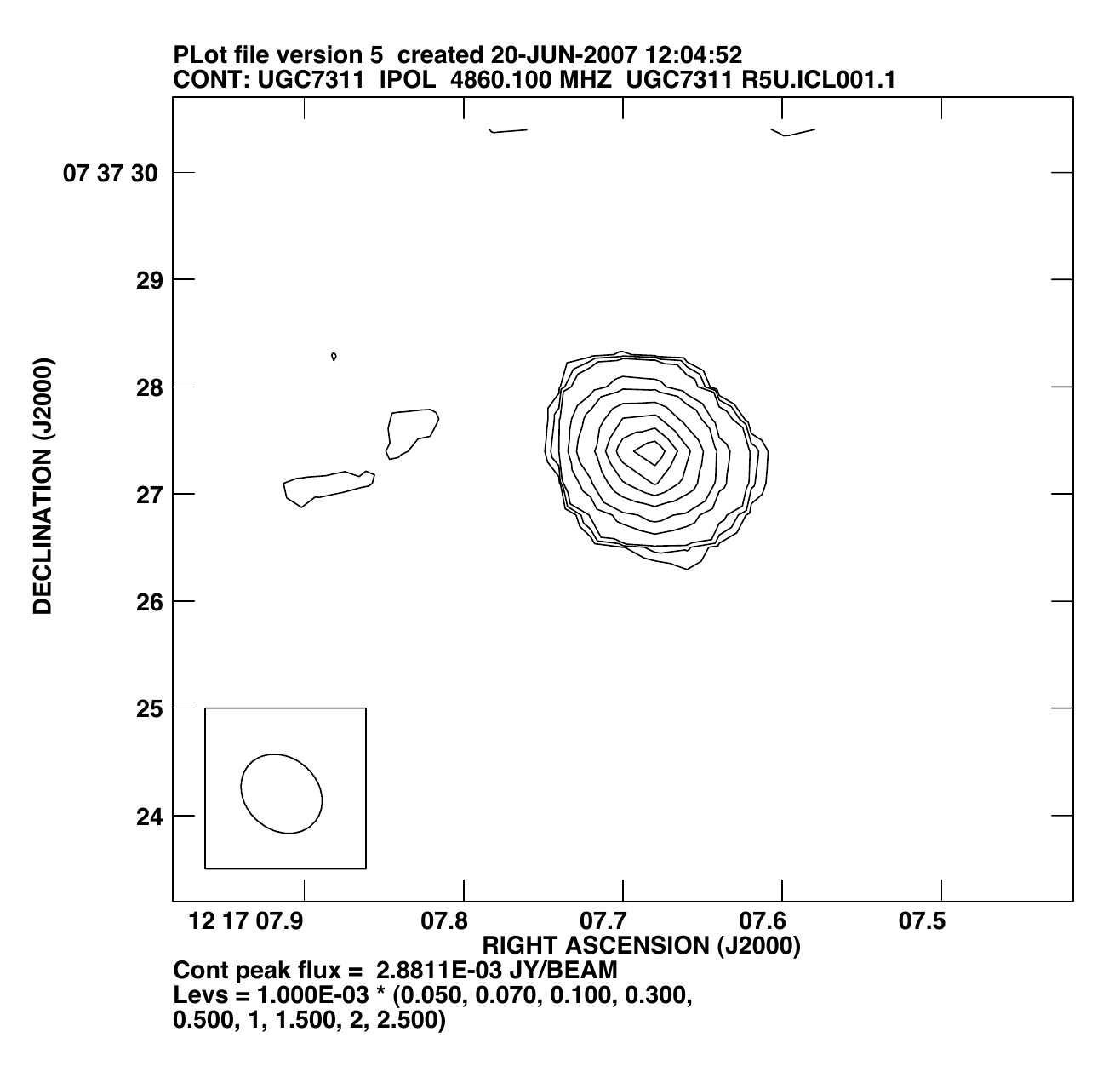}
\caption{\small (continued) }
\end{figure}
\addtocounter{figure}{-1}
\begin{figure}
\includegraphics[width=9.2cm]{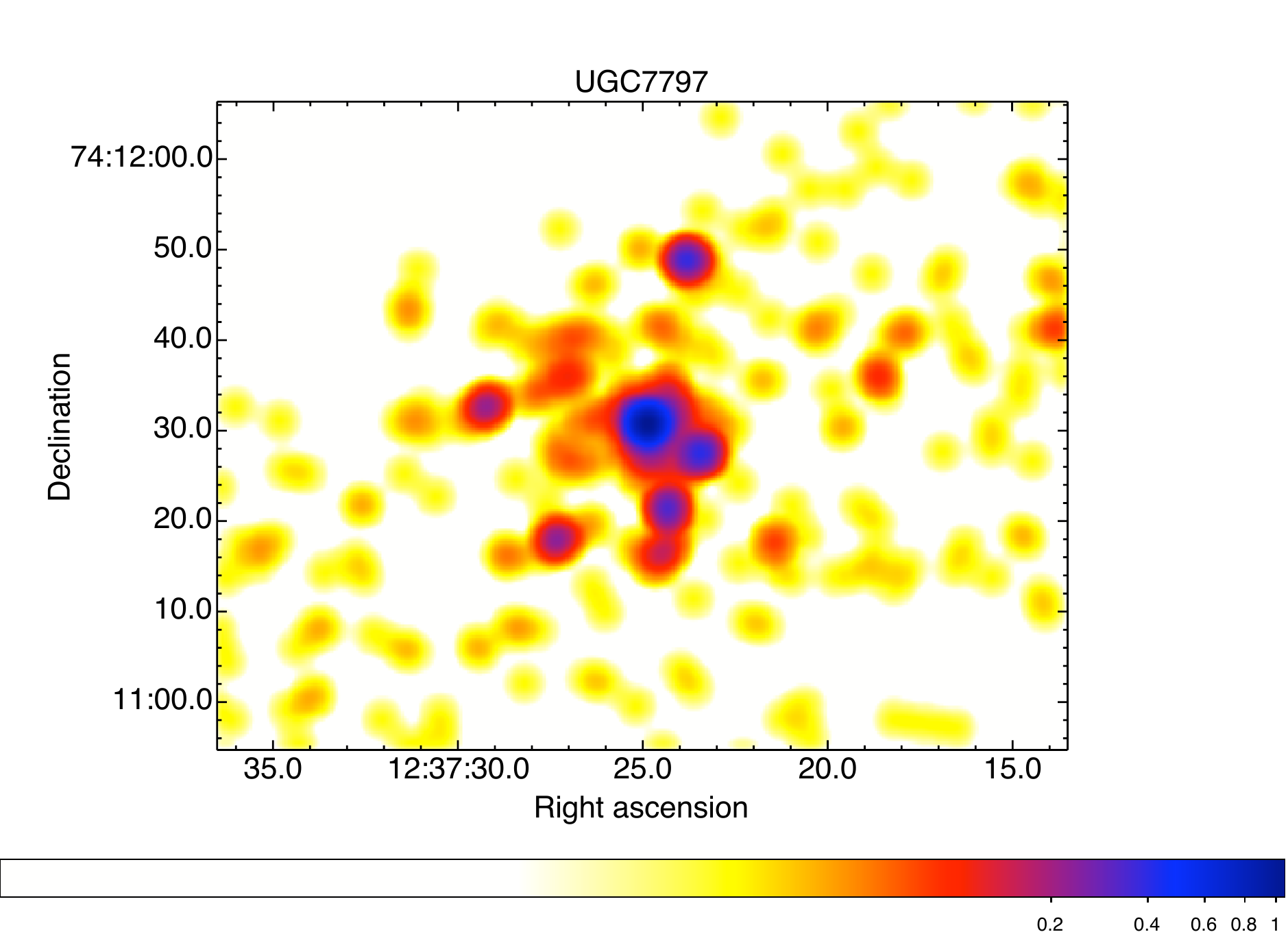}
\includegraphics[width=7.6cm]{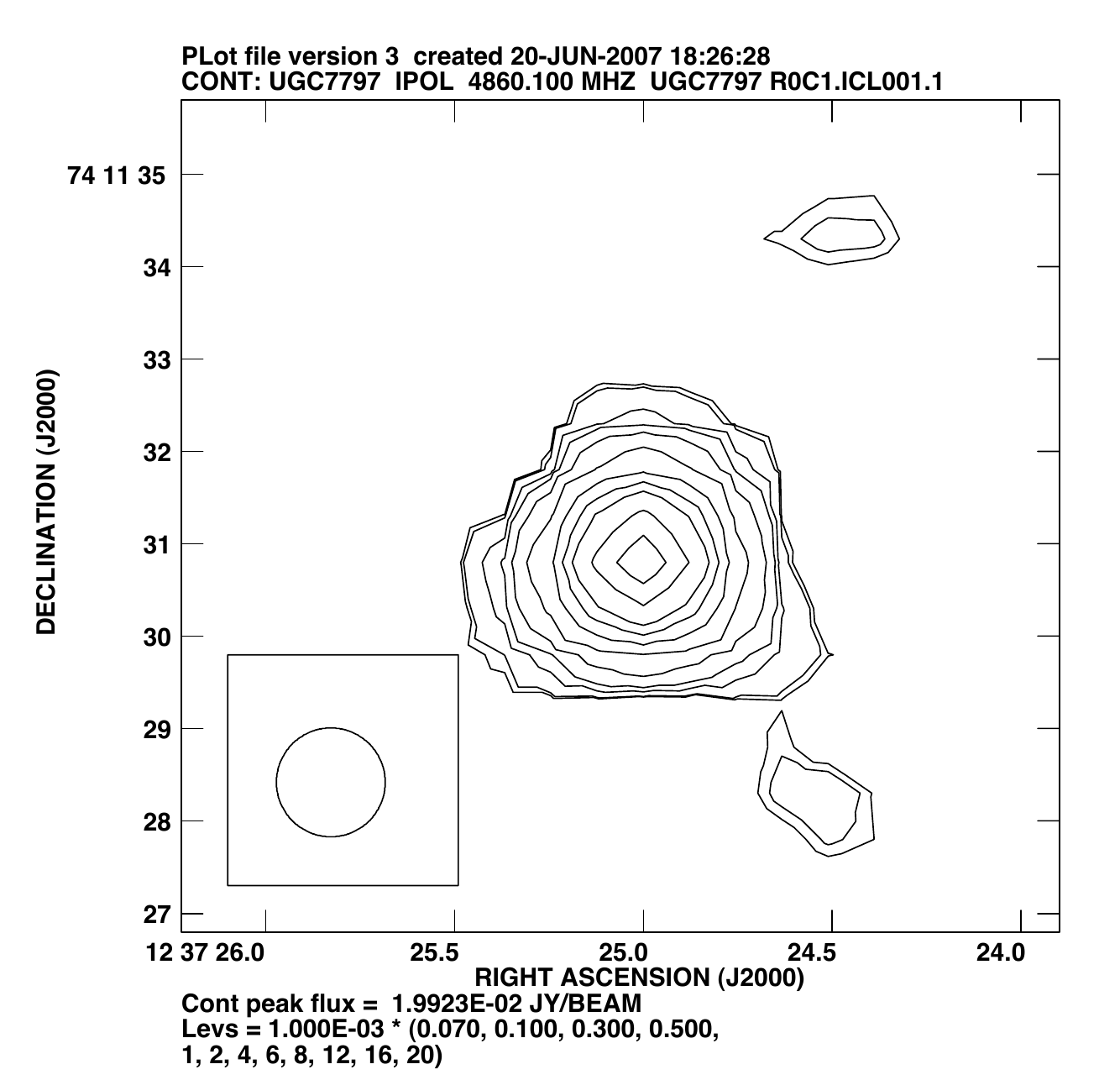}
\includegraphics[width=9.2cm]{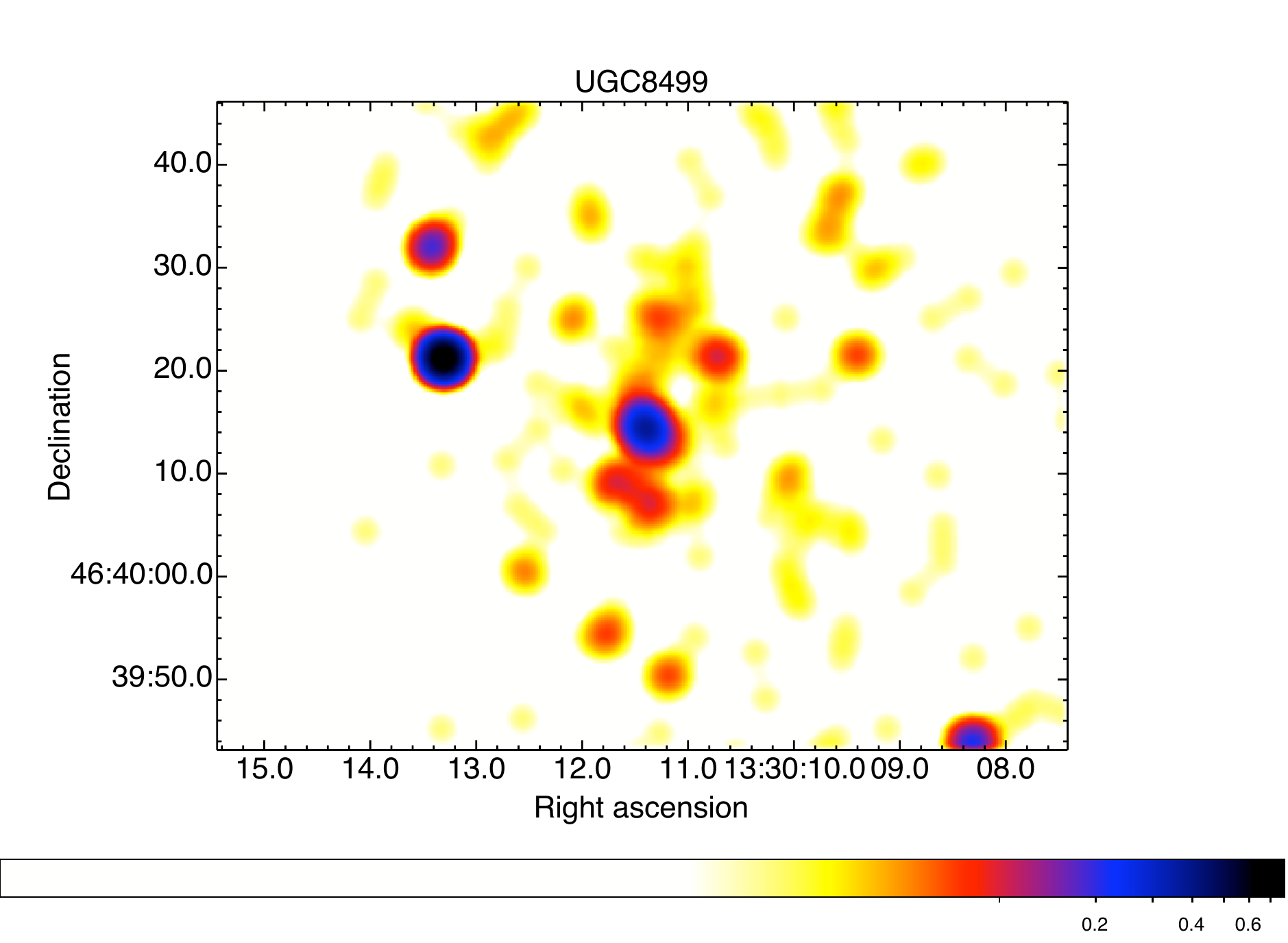}
\includegraphics[width=7.6cm]{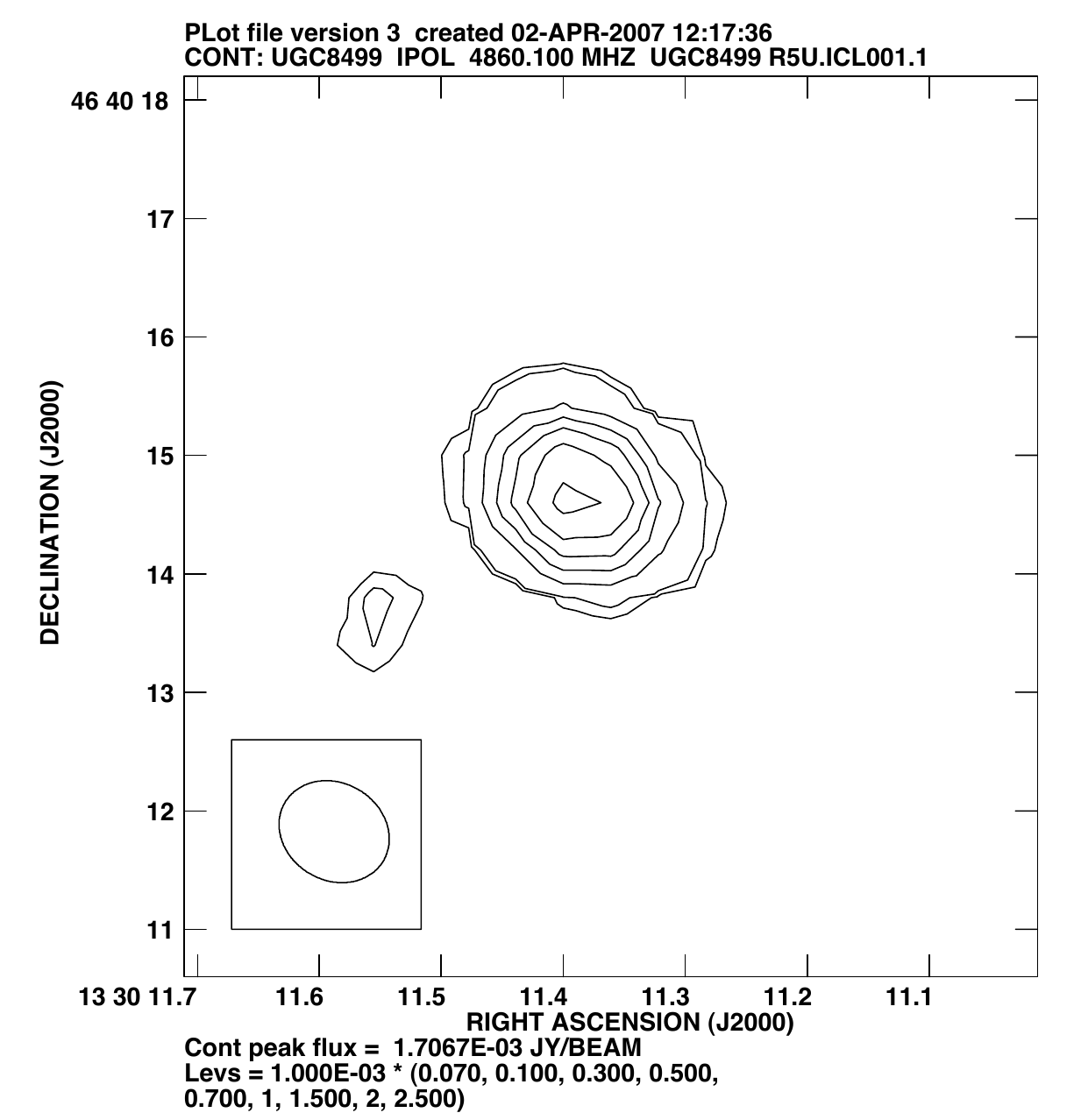}
\caption{\small (continued) } 
\end{figure}
\addtocounter{figure}{-1}
\begin{figure}
\includegraphics[width=9.2cm]{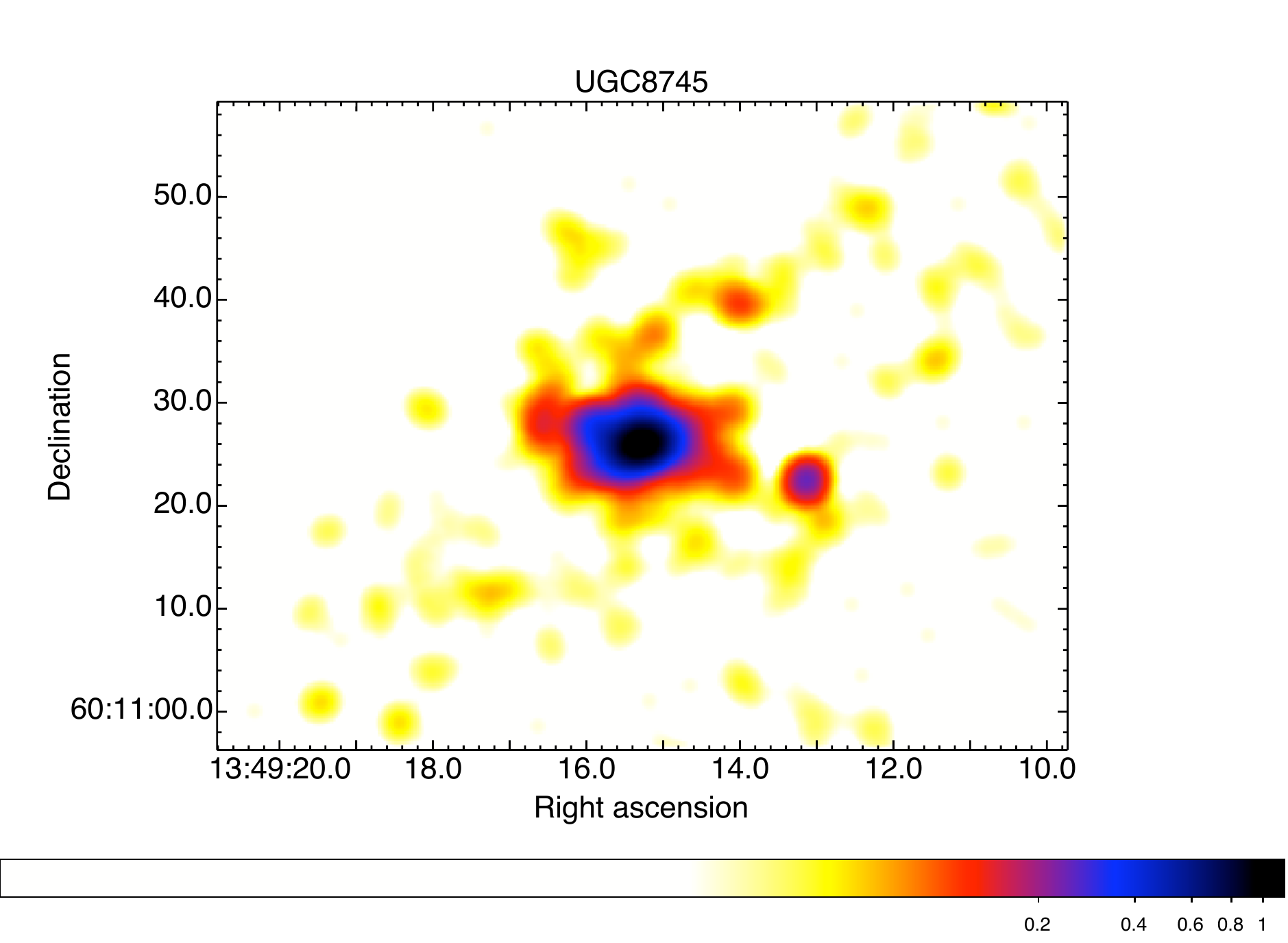}
\includegraphics[width=6.6cm]{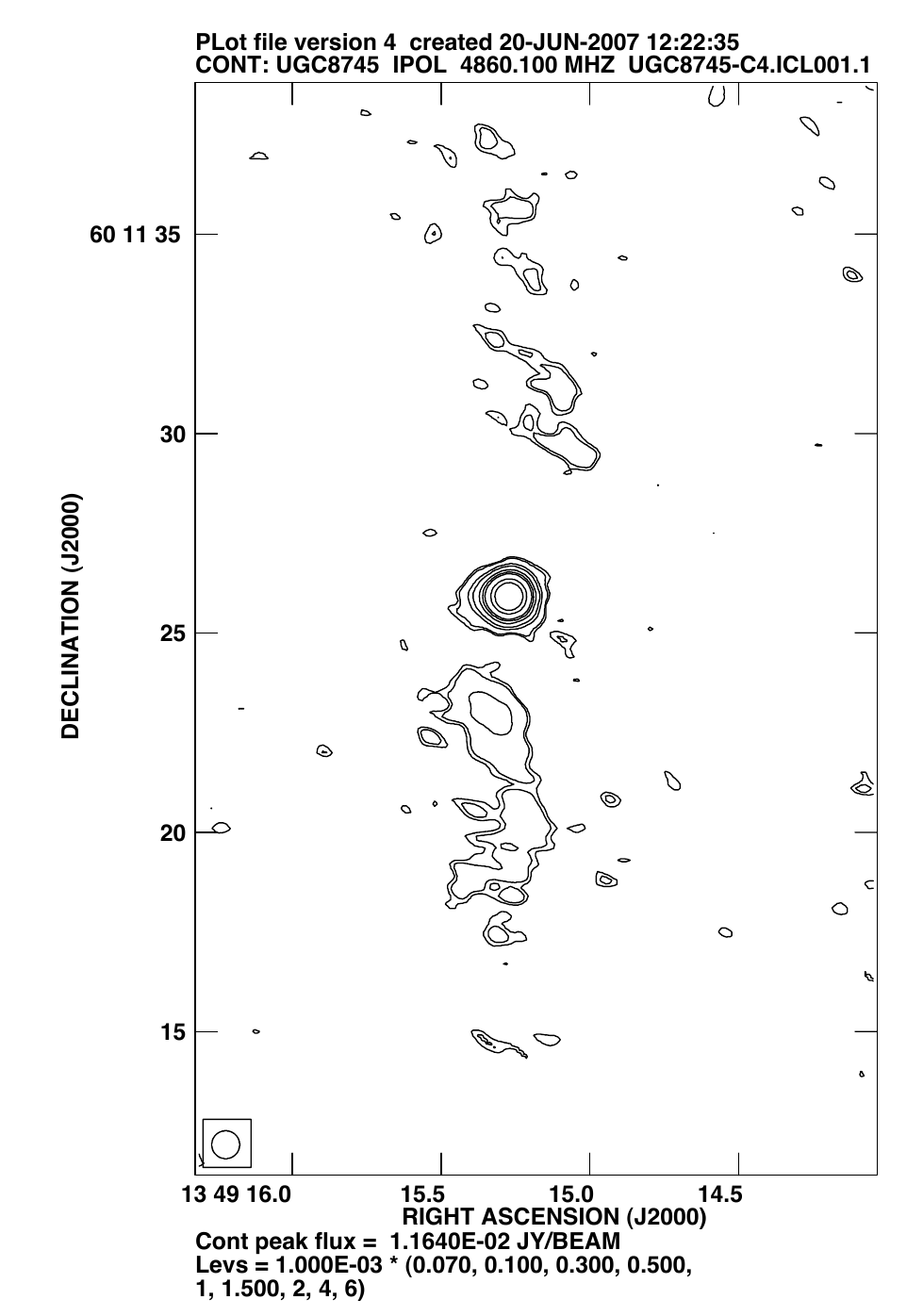}
\includegraphics[width=9.2cm]{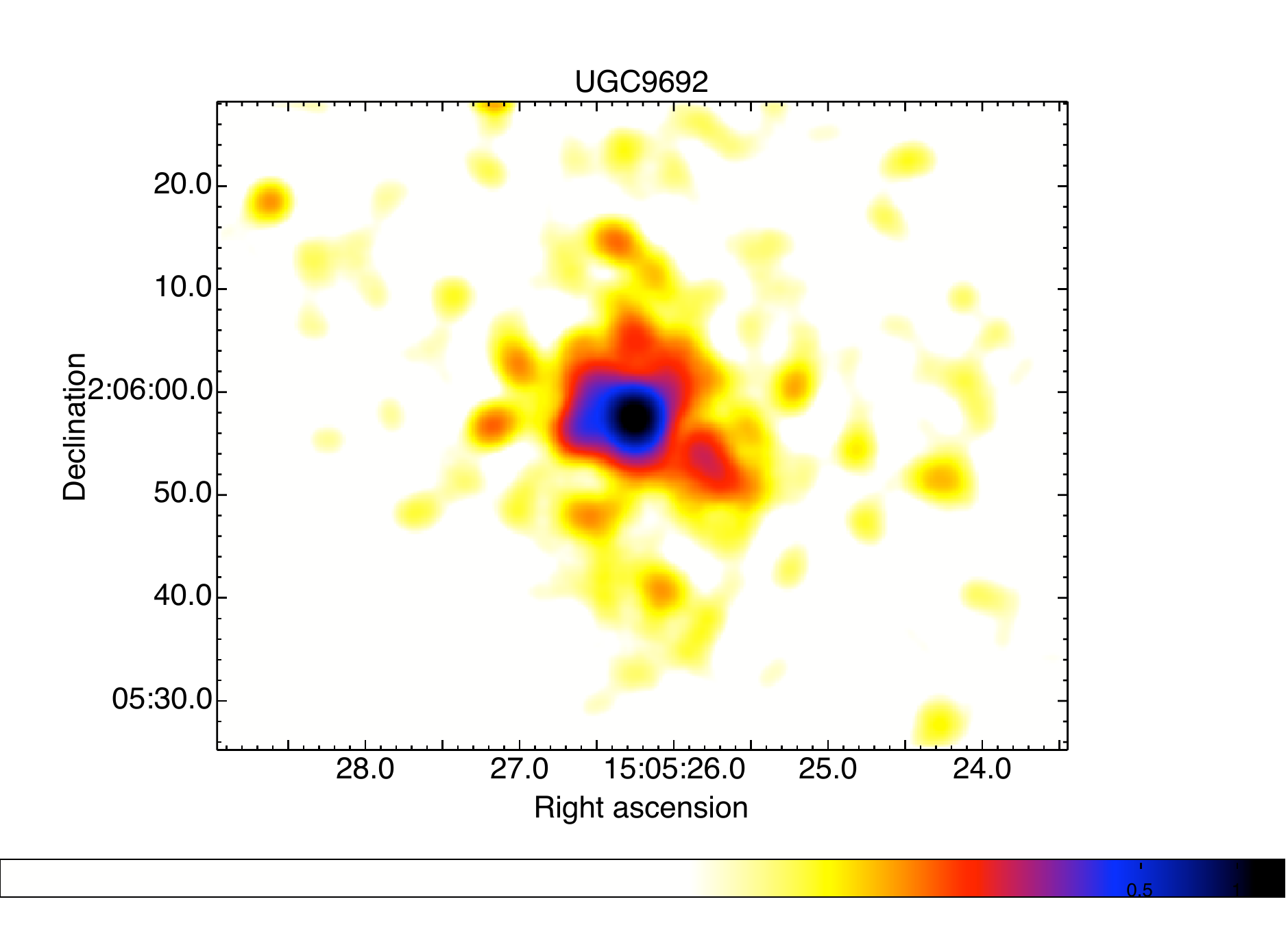}
\includegraphics[width=7.6cm]{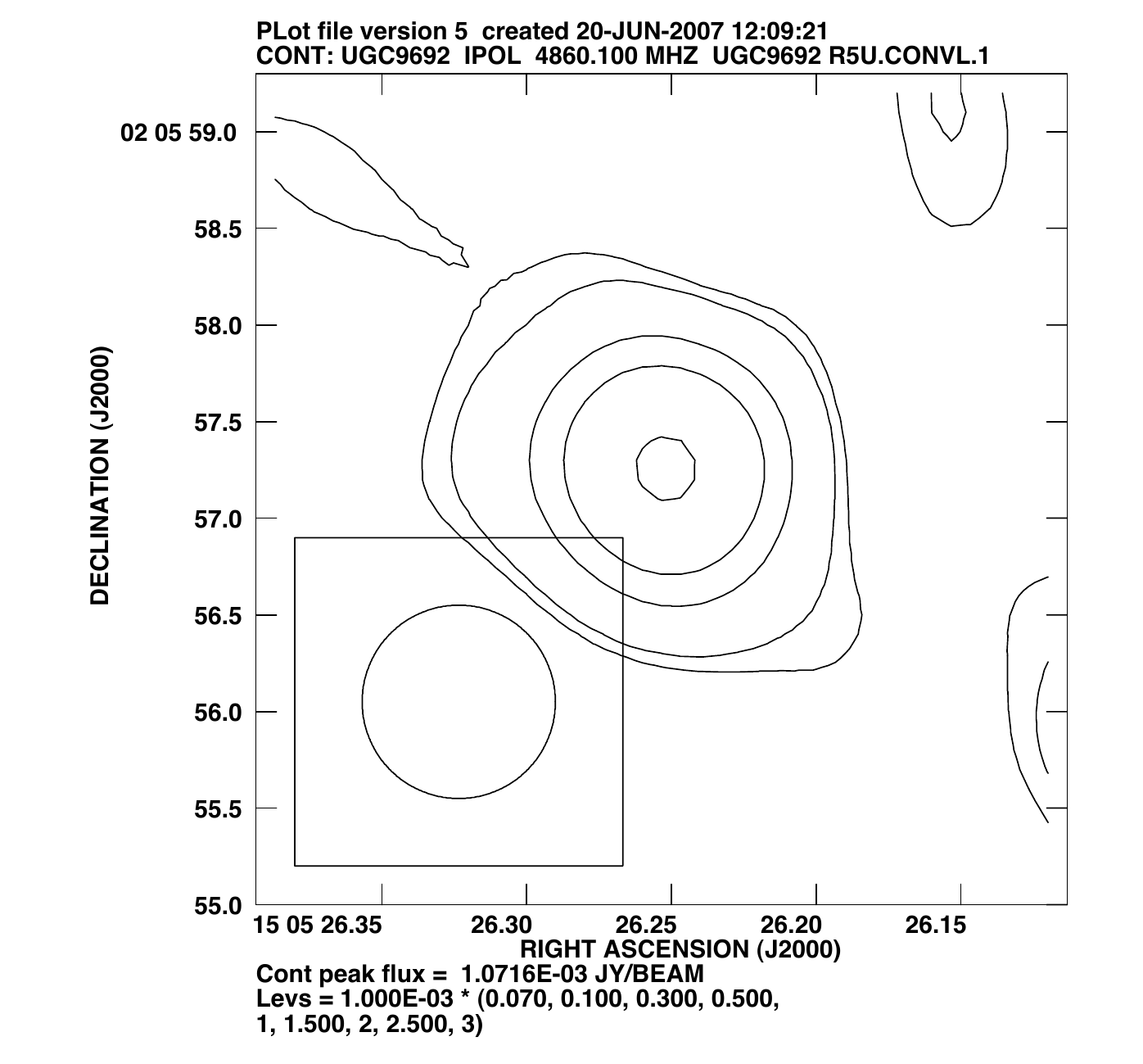}
\caption{\small (continued) }
\end{figure}
\addtocounter{figure}{-1}
\begin{figure}
\includegraphics[width=9.2cm]{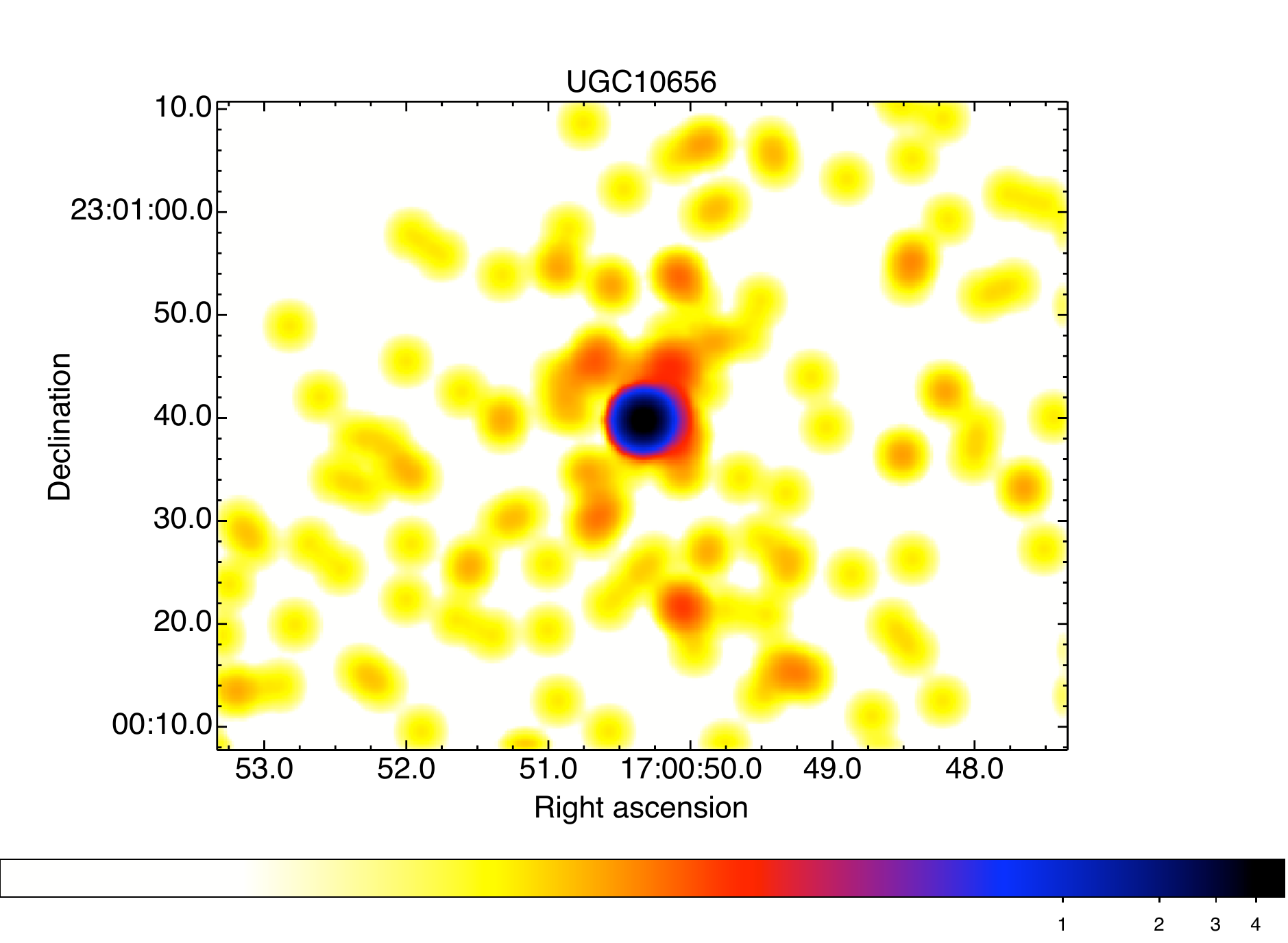}
\includegraphics[width=7.6cm]{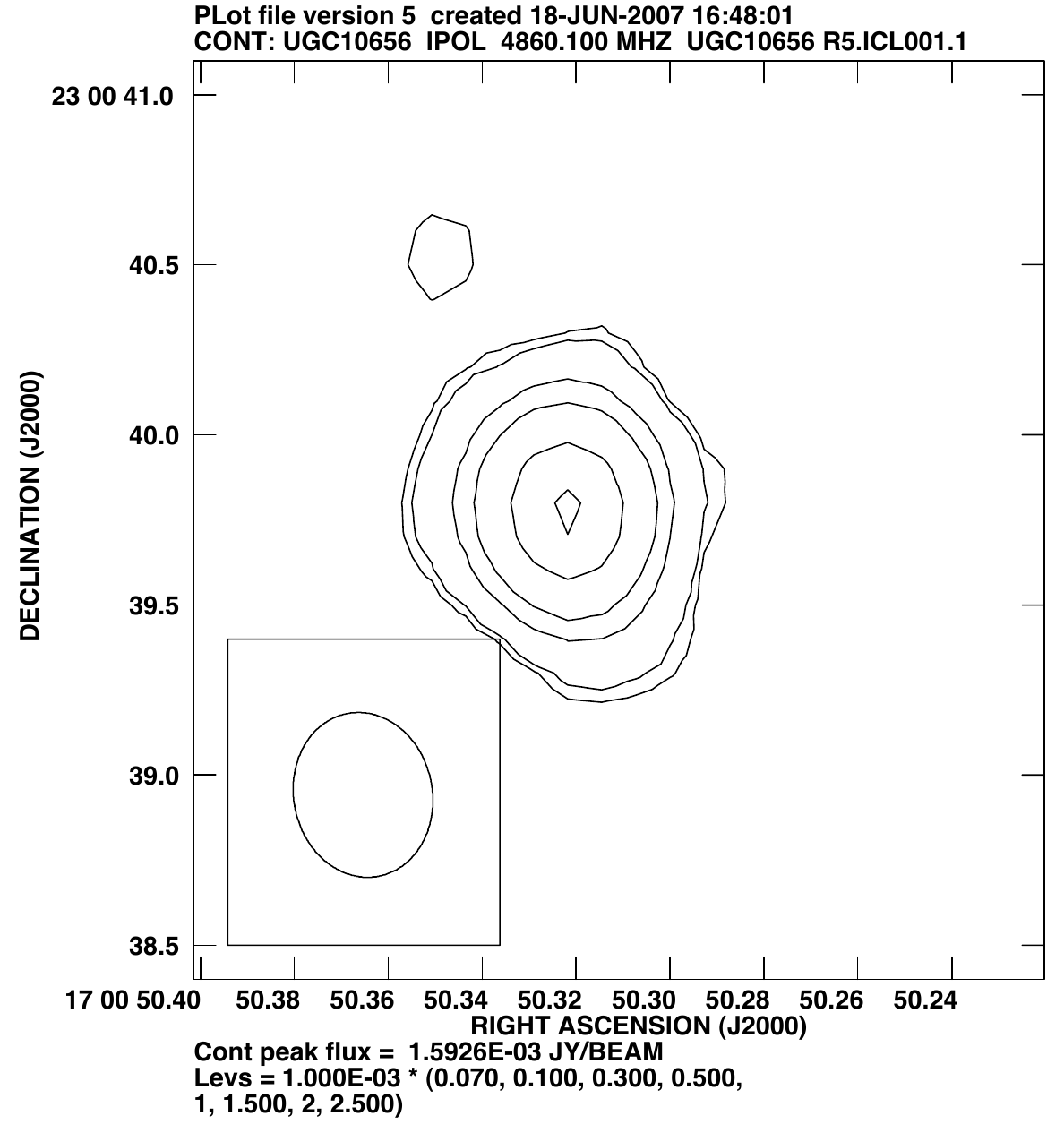}
\caption{\small (continued)} 
\end{figure}

\begin{deluxetable}{lcccllcllclc}
\rotate
\tabletypesize{\scriptsize}
\tablecaption{Radio and X-ray estimates for the 13 UGC galaxies}
\tablewidth{0pt}
\tablehead{
\colhead{Source} & \colhead{ObsID} & \colhead{$z$} & \colhead{$F_{r}^{W91}$}  & \colhead{$F_{r}$} & \colhead{SBP} & \colhead{$N_{H, Gal}$} 
& \colhead{Count rate} & \colhead{$\Gamma$ (error)} & \colhead{$F_{x}$ (error)} & \colhead{$\chi^{2}$/dof} & \colhead{HR (error)} \\
\colhead{(1)}&\colhead{(2)}&\colhead{(3)}&\colhead{(4)}&\colhead{(5)}&\colhead{(6)}&\colhead{(7)}&\colhead{(8)}&\colhead{(9)}&\colhead{(10)}& \colhead{(11)} & \colhead{(12)}}
\startdata
UGC\,0968 & 6778   & 0.00793  & 1.4 &  1.5   &  Core  & 0.047  &1.1E$-$02& 1.7 ($-$0.2, 0.3)           & $<$7.9E$-$15 ($-$1.0E$-$15, 1.1E$-$15)  &... & ... \\
UGC\,5959 & 6779   &  0.00471 & 5.0 &  ...         & PLaw  & 0.022  &4.5E$-$02& 1.7 ($-$0.2, 0.3)           & 2.6E$-$13 ($-$1.8E$-$14, 3.1E$-$14)  &38/21& 0.41 (0.03)\\
UGC\,6860 & 6780   &  0.00420 & 1.0 &  1.9      &  Inter & 0.016  &3.4E$-$02&  1.7 ($-$0.2, 0.3)         & 1.7E$-$13 ($-$3.4E$-$15, 8.3E$-$15)    &30/14& 0.69 (0.03)\\ 
UGC\,6946 & 6781   & 0.00346  & 83. &  302.      & PLaw   & 0.012  &1.4E+00&  1.75 ($-$0.02, 0.02) & 1.2E$-$11 ($-$3.6E$-$12, 2.1E$-$12)   &172/161& 0.32 (0.007)\\
UGC\,6985 & 6782   &  0.00310 & 1.4  & $<$0.12 & PLaw   & 0.019  &5.7E$-$03&  1.7 ($-$0.2, 0.3)        & $<$1.1E$-$14 ($-$4.2E$-$16, 5.1E$-$16)   &...& ... \\
UGC\,7005 & 6783   & 0.00482  & 2.9 &   3.3      & Inter   & 0.018  &1.0E$-$02& 1.7 ($-$0.2, 0.3)        & 3.6E$-$14 ($-$2.4E$-$15, 2.8E$-$15)  &39/4 & -0.25 (0.09)\\
UGC\,7311 & 6784   &  0.00790 & 1.9 &     3.0       & PLaw  & 0.015  &1.4E$-$02&  1.7 ($-$0.2, 0.3)          &1.2E$-$13 ($-$1.9E$-$15, 1.4E$-$15)  &19/5& -0.09 (0.07)\\
UGC\,7797 & 6785   &  0.00660 & 21. &   20.4       & Core  & 0.020  &7.7E$-$03&  1.7 ($-$0.2, 0.3)           & $<$2.1E$-$14 ($-$1.2E$-$15, 1.2E$-$15)  &...& ... \\
UGC\,8499 & 6786   &  0.00844 & 1.5 &   1.8       & Core$^{\star}$  & 0.017  &5.2E$-$03&  1.7 ($-$0.2, 0.3) & $<$6.2E$-$15 ($-$2.5E$-$16, 1.9E$-$16) &...& ... \\
UGC\,8745 & 6787   &   0.00585& 20. &  11.6       & Core   & 0.017  &1.2E$-$02&  1.7 ($-$0.2, 0.3)           & $<$1.5E$-$14 ($-$3.7E$-$16, 4.1E$-$16)   &...& ... \\
UGC\,9692 & 6788   &  0.00453 & 2.0 &   1.1       & PLaw  & 0.041  &7.4E$-$03&  1.7 ($-$0.2, 0.3)           & 2.1E$-$14 ($-$5.3E$-$16, 5.9E$-$16)   &16/6& 0.49 (0.12)\\
UGC\,10656& 6789   &  0.00930 & 1.2 &   1.6     & PLaw  & 0.048&1.7E$-$02& 1.7 ($-$0.2, 0.3)               &1.0E$-$13 ($-$6.1E$-$15, 9.8E$-$15)   &9/8& 0.40 (0.06) \\
UGC\,12759&6790    &  0.00570 & 2.7 &   1.1      & PLaw  & 0.052&9.8E$-$03&  1.7 ($-$0.2, 0.3)              & 2.7E$-$14 ($-$5.4E$-$15, 6.3E$-$15)   &19/2& 0.88 (0.06)\\ 
\enddata
\tablecomments{Col.\,1: Source name. Col.\,2: {\it Chandra} Observation ID. Col.\,3: Source redshift {from NED\footnote{NASA/IPAC Extragalactic Database}}. Col.\,4: 4.9 GHz flux density in mJy from \citet{wrobel91b,wrobel91a}. Col.\,5: Flux density at 4.9 GHz in mJy from our VLA observations. Flux densities for UGC\,0968 and UGC\,12759 were obtained from Henrique Schmitt (private communication). {The rms noise in the radio maps is typically $\sim0.05$ mJy\,beam$^{-1}$. The $3\sigma$ upper limit for the UGC\,6985 flux density is quoted.} Col.\,6: HST nuclear surface brightness profile from Paper I: Core = core, PLaw = power-law, Inter = Intermediate. ${\star}$ UGC\,8499 has recently been reclassified as a core galaxy (cf. Section 4.1)  Col.\,7: Galactic hydrogen column density in units of $10^{22}$\,cm$^{-2}$ {obtained through the Chandra COLDEN tool\footnote{http://cxc.harvard.edu/toolkit/colden.jsp}}. Col.\,8: X-ray count rate in counts sec$^{-1}$. Col.\,9: Photon index with error. For UGC\,6946, the XSPEC spectral fitting errors are quoted at the 90\% confidence level for one parameter of interest. For all the other sources, ``error'' indicates that the spectral fitting was carried out while keeping the photon index fixed to 1.5 and 2.0 (cf. Section 3). Col.\,10: Unabsorbed 0.5$-$5 keV flux density in erg~cm$^{-2}$~sec$^{-1}$ and error at the 90\% confidence level for one parameter of interest. Col.~11: $\chi^{2}$ by degrees of freedom. Cols.~12: Hardness ratio and error.}
\label{tabsample}
\end{deluxetable}

\section{Results}
Nuclear X-ray emission is detected in eight sources while radio emission is detected in all but one of them, {\it viz.}, UGC\,6985 (see Appendix A for notes on individual sources).
{We note that the offset between the peak of the X-ray emission and the radio cores (measured directly from the images) is typically around $0\farcs1$. Only UGC\,7797 has an X-ray-radio peak offset of $0\farcs2$, while UGC\,8499 has an offset of about $0\farcs6$.
Looking at the HST WFPC2\footnote{Wide Field Planetary Camera 2} archival images for all sources without radio maps, we also find that the typical offset between the peak of the X-ray and optical emission is about $0\farcs4$ or less. Only UGC\,6985 has an X-ray-optical peak offset of about $0\farcs7$. These offsets are consistent with the typical astrometric uncertainty of {\it Chandra} and {HST} images. 
It is unlikely that the X-ray emission is due to stellar sources.
The small positional offsets of the X-ray cores from the radio and optical nuclei ($0\farcs1$ translates to linear scales of 7 to 18 parsec at the distance of the nearest and farthest sample galaxy, respectively), support the conclusion that they are AGN. Furthermore, the X-ray sources have luminosities $>10^{38}$ erg~sec$^{-1}$, compared with typical luminosities 
$<10^{38}$ erg~sec$^{-1}$ for X-ray binaries \citep{Miller05}.}

The X-ray emission seems to be extended in some galaxies (e.g., UGC\,0968, UGC\,8745, see Figure \ref{figONE}). In two galaxies, {\it viz.,} UGC\,9692 and UGC\,7005, the extension in the X-ray emission seems to have a similar position angle as the radio emission. 
However, the extension of the radio structure is on much smaller spatial scales (few arcsec) than that in the X-rays (tens of arcsec). 
Therefore, it is difficult to make a case for jet-ISM interaction and shock-heating giving rise to the X-ray emission in these galaxies. The X-ray extension in UGC\,8745 is clearly in a direction perpendicular to the radio jet, but along the dust lane. The X-ray emitting gas in this source could therefore be associated with the galactic disk plane. 

Since the small number of X-ray counts (typically $<$100) precluded an analysis with sophisticated spectral models, only a simple absorbed power-law model with absorption set to the Galactic value, and a fixed photon index of 1.7, could be fitted to the X-ray spectra of all but one source (UGC\,6946). The 0.5$-$5 keV X-ray flux density varies between $5\%-10\%$ when the fixed photon index {is changed to} 1.5 and 2.0. The results from the spectral fitting analysis are presented in Table \ref{tabsample}. 

We estimated the hardness ratios (HR) {in the source nuclei} using the relation, HR = (S$-$H)/(S$+$H), where S refers to the count rate in the soft energy band, 0.5$-$2~keV, and H refers to the count rate in the hard energy band, 2$-$10~keV (Table \ref{tabsample}). A source with a small HR value is expected to be heavily obscured, with all photons being emitted at relatively high energies. The counts in the S and H bands were obtained {from} the same regions that were used for the nuclear spectral extraction for flux estimation in XSPEC. We used the Chandra Education Analysis Tools as implemented in the Virtual Observatory option in DS9, to estimate the counts along with the error in counts. The hardness ratios and errors were estimated following equations 2 and 3 of \citet{Park06}. We find that the HR values fall typically between 0.3 and 0.9. Much lower values are measured for the ``power-law'' source UGC\,7311 ($-0.09\pm$0.07) and for the ``intermediate" source UGC\,7005 ($-0.25\pm$0.09), which are likely to be heavily absorbed.

We discuss {below} the multiwavelength (radio-optical-Xray) properties of the entire sample of 51 radio-selected early-type galaxies, that has been presented in Papers I $-$ III and the current paper. 

\begin{figure}
\centering{
\includegraphics[width=10.5cm]{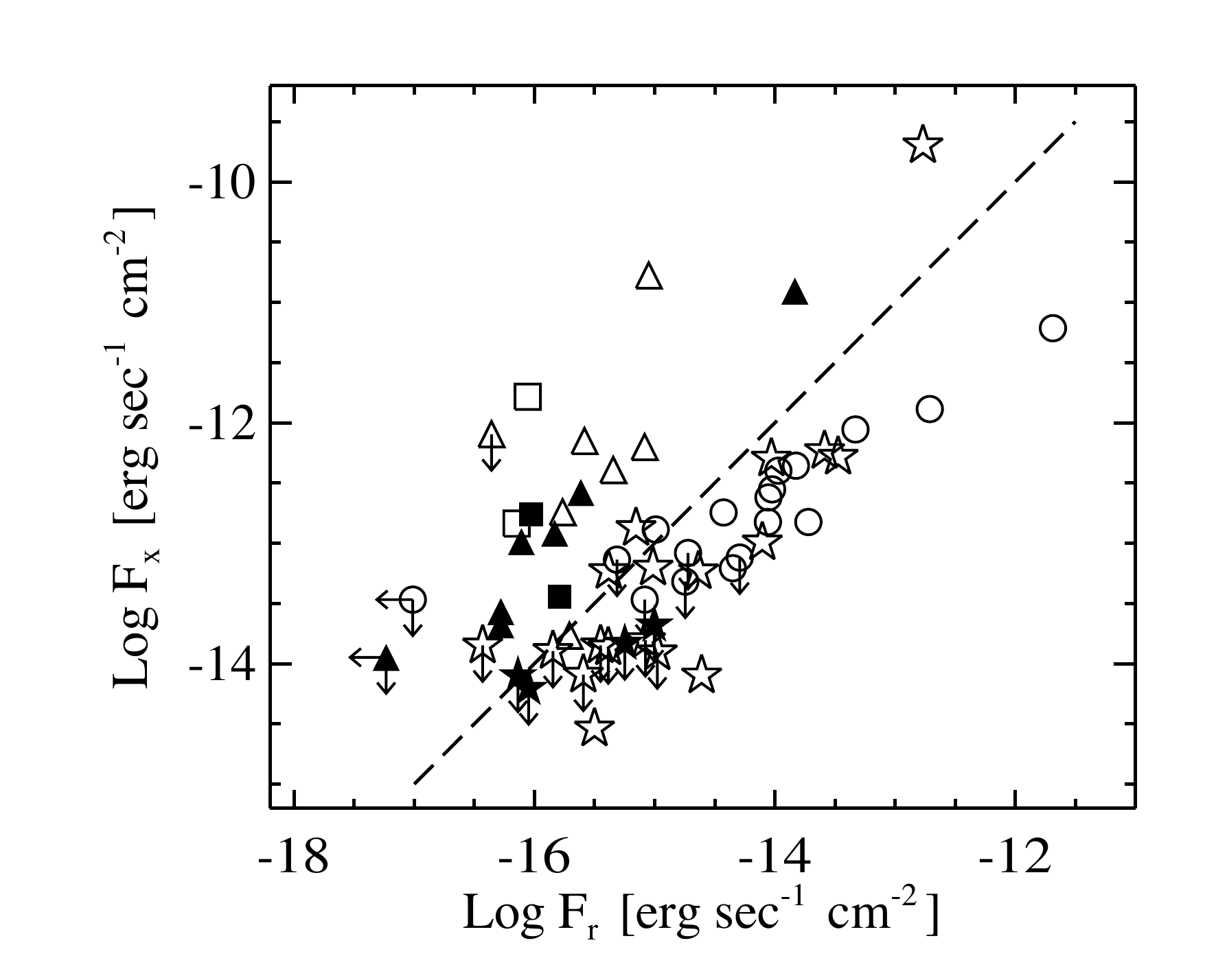}
\includegraphics[width=10.5cm]{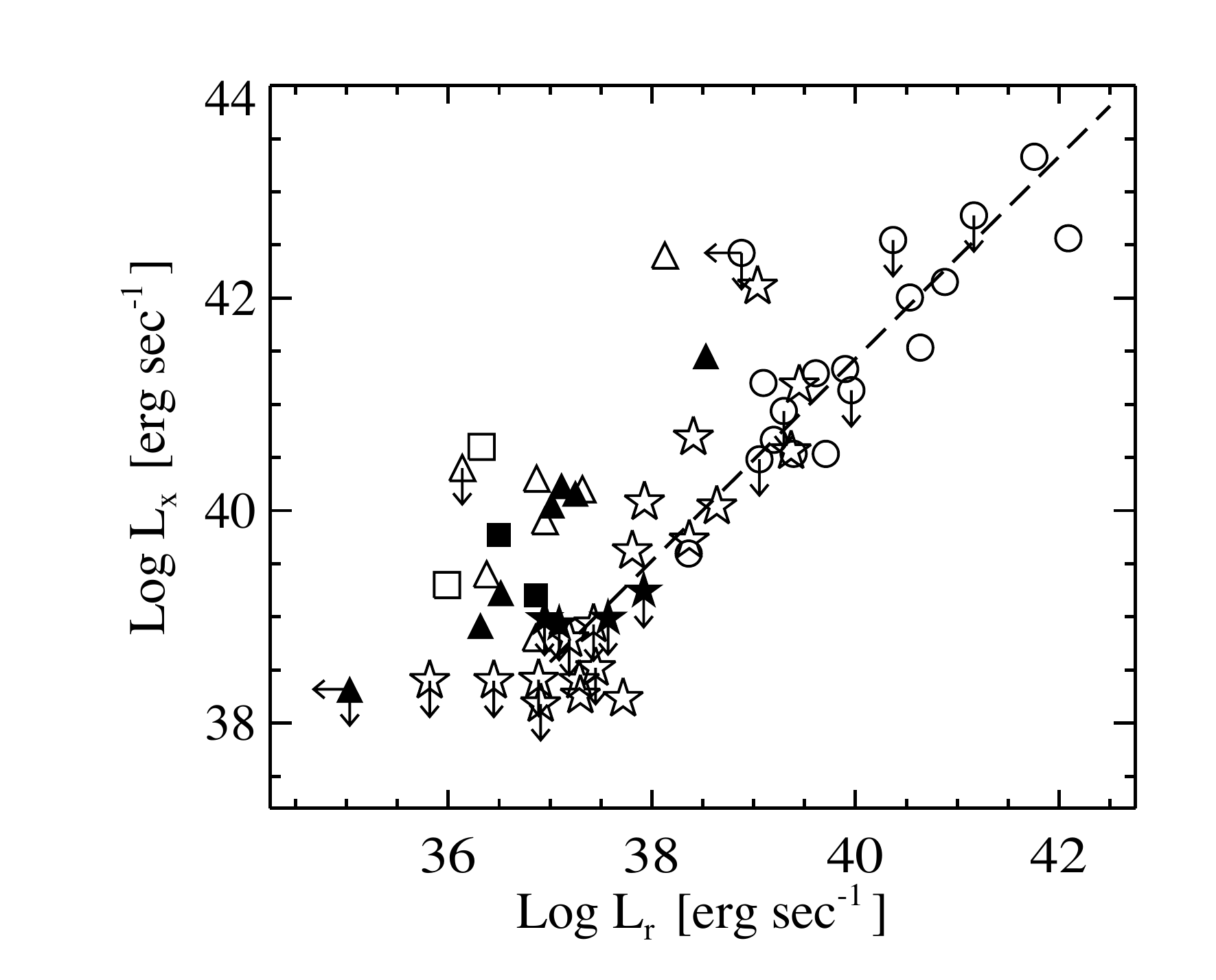}}
\caption{\small The 4.9 GHz radio versus the 0.5$-$5 keV X-ray flux density (top panel) and luminosity (bottom panel) for the entire sample. The stars, triangles and squares denote ``core'', ``power-law'' and ``intermediate'' galaxies, respectively. The circles denote FRI radio galaxies from \citet{Balmaverde06}. The open symbols represent sources from Papers II \& III, while the filled symbols indicate the 13 new sources presented in this paper. The dashed line in the top panel is a constant X-ray-radio ratio (=100) line, while that in the bottom panel is the best linear fit to the FRI radio galaxy data alone, which has a slope of 0.95.}
\label{figTWO}
\end{figure}

\subsection{Dependence of the X-ray \& Radio Nuclear Emission on the Optical Surface Brightness Profile}

Plotted in the top panel of Figure \ref{figTWO} are the 0.5$-$5 keV X-ray versus the 4.9 GHz radio flux densities for the entire sample. 
{The radio flux densities quoted in mJy in Table \ref{tabsample} were converted to erg~cm$^{-2}$~sec$^{-1}$ in Figure \ref{figTWO}, using the central frequency of 4.86 GHz.}
The bottom panel of Figure \ref{figTWO} displays the X-ray versus the radio luminosities. 
The dashed line in the flux density panel is a constant ratio (=100) line, while the line in the luminosity panel indicates a linear fit to the FRI radio galaxy data alone, {following} the least absolute deviation method, which is implemented as LADFIT in IDL. {This line has a slope of 0.95, with the mean of the absolute deviation being = 0.41.}

We note here that the X-ray flux densities for the FRI radio galaxies {in Figure \ref{figTWO}} are from the 0.5$-$5 keV energy band \citep{Balmaverde06}, while the X-ray flux densities for the ``core'', ``power-law'' and ``intermediate'' galaxies from Papers II and III, are from the 2$-$10 keV energy band. However, we find that the 2$-$10 keV X-ray flux estimates for the 13 new UGC galaxies differ by only $\sim$10\% from the plotted 0.5$-$5 keV flux values for our choice of the photon spectral index (=1.7). Therefore, the choice of the energy-band does not affect any of our findings discussed below. 

In the radio-X-ray plane, the three new ``core'' galaxies lie on the low luminosity end of the well established FRI radio galaxy correlation (bottom panel of Figure \ref{figTWO}), similar to the other ``core'' galaxies. Seven of the eight new ``power-law'' galaxies lie well above the FRI correlation while the two new ``intermediate'' galaxies appear to lie largely in the region occupied by the ``power-law'' galaxies, again consistent with the rest of the ``power-law/intermediate'' galaxies. It is evident that more ``intermediate'' galaxies are required to confirm this {trend}. {It is worth noting that} the two ``power-law'' galaxies with low HR are likely to be obscured and hence their intrinsic X-ray fluxes must be higher than those quoted/plotted here. {Additionally,} UGC\,8499 was originally classified as a ``power-law'' galaxy in Paper I, but has subsequently been reclassified as a ``core'' galaxy by \citet{Lauer07}, {since} unduly conservative resolution limits were assumed for the deconvolved HST WFPC2 images by \citet{Rest01}.
We {did not} detect nuclear emission {either} in the radio {or} in the X-rays in UGC\,6985. Moreover, no optical emission lines are visible in the spectra of this source \citep{ho97b}. Therefore, there is no evidence that this galaxy has an {AGN}.

\subsubsection{Broad-band Spectral Indices}
\begin{figure}
\centering{
\includegraphics[width=8.194cm]{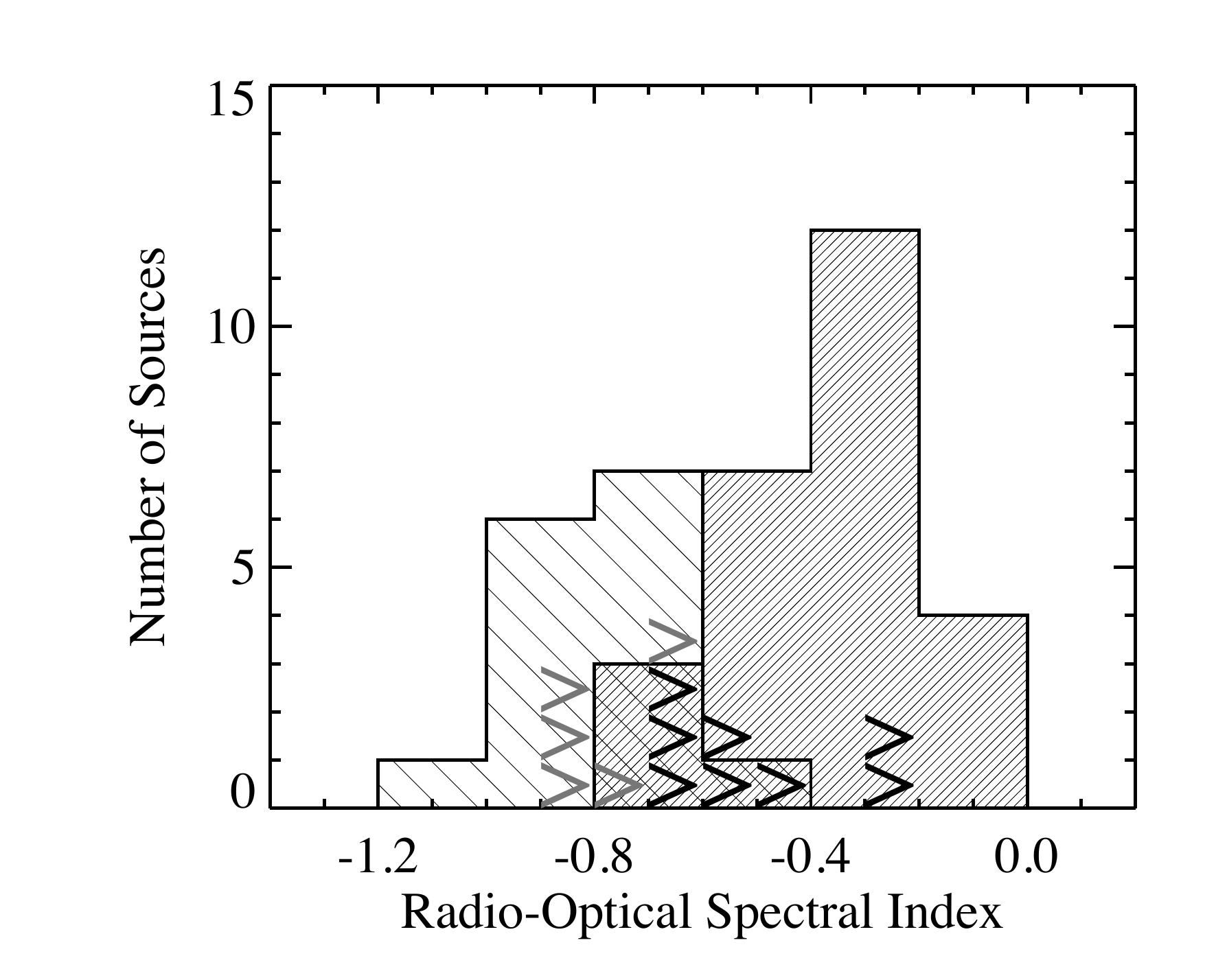}
\includegraphics[width=8.194cm]{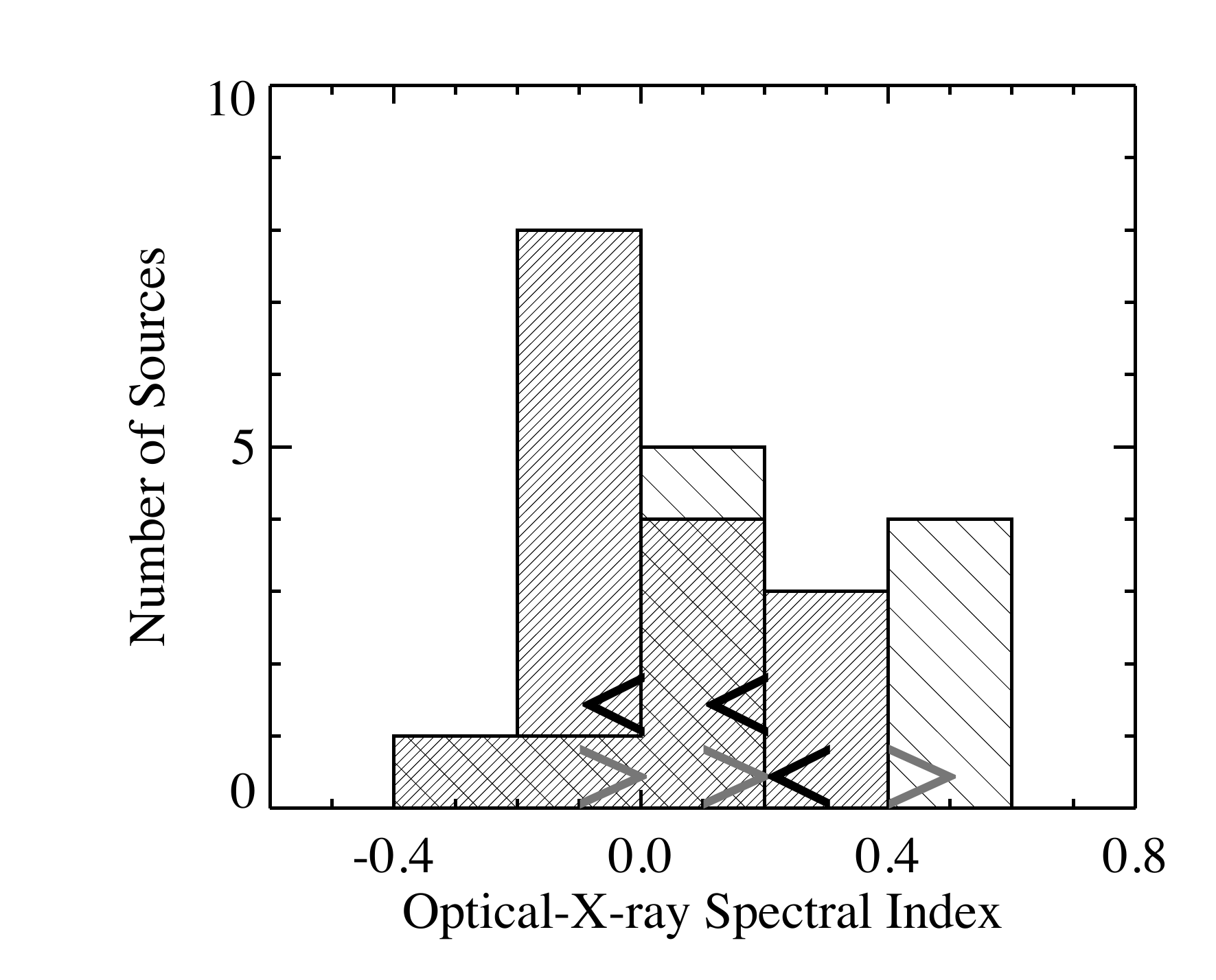}
\includegraphics[width=8.194cm]{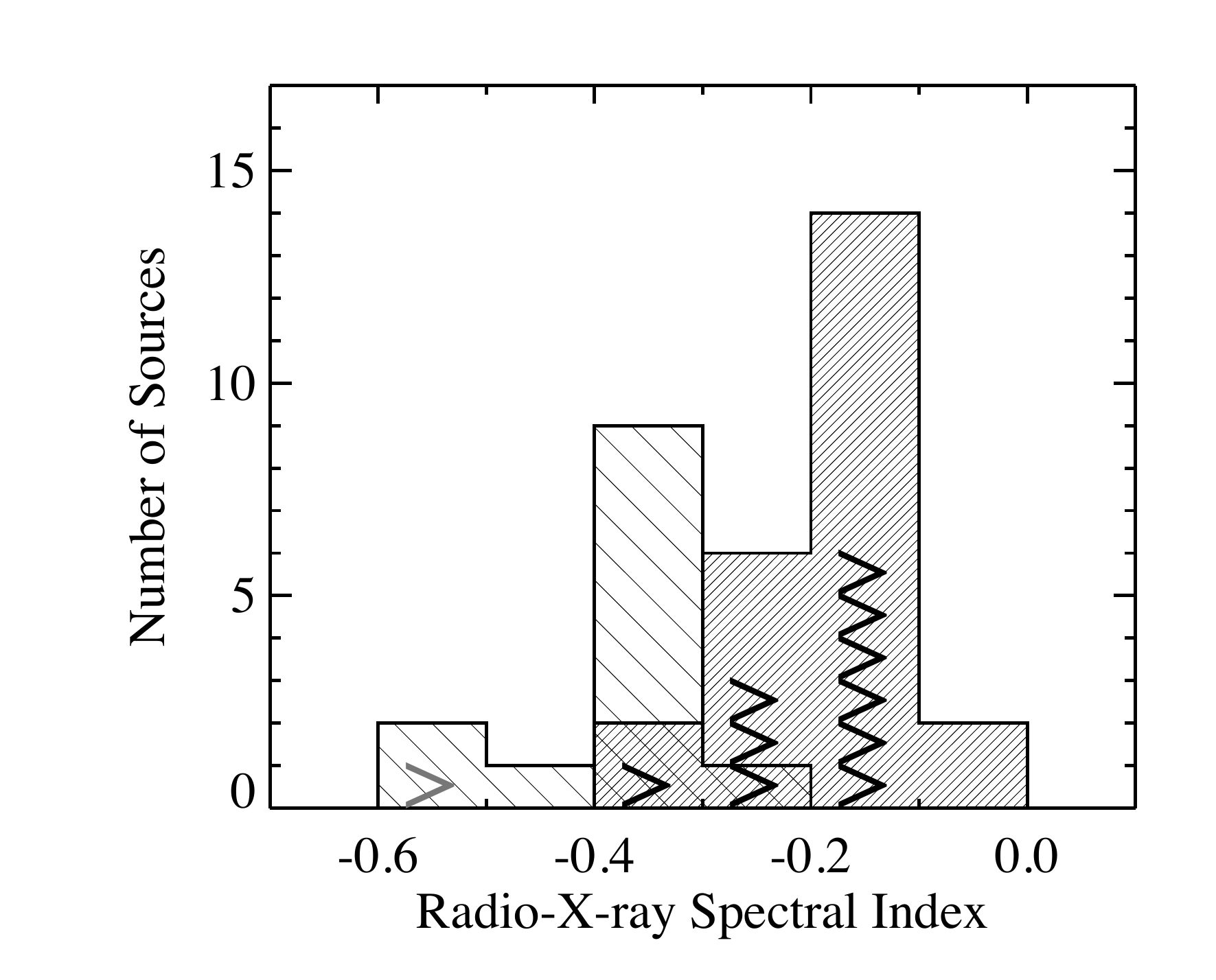}}
\caption{\small Two-point spectral indices for the ``core'' (darker shaded regions) and ``power-law'' (hashed regions with +45$\degr$ lines) galaxies. {Upper and lower limits are indicated by ``$<$'' and ``$>$'', respectively: they are black for ``core'' and gray for ``power-law'' galaxies. ``Power-law'' galaxies have flatter (or more inverted) $\alpha_{ro}$ and $\alpha_{rx}$ compared to ``core'' galaxies (top left and bottom panel). $\alpha_{ox}$ are not statistically different between the ``core'' and ``power-law'' galaxies (top right panel).}}
\label{figTHREE}
\end{figure}

Using the HST optical, {Chandra X-ray and VLA radio} core flux densities reported in \citet{BalmaverdeCapetti06,CapettiBalmaverde06}, we have estimated the radio-to-optical ($\alpha_{ro}$), optical-to-Xray ($\alpha_{ox}$) and radio-to-X-ray ($\alpha_{rx}$) spectral indices for the entire sample (Table \ref{tabalpha}). We note that the optical data has been obtained through different HST filters and cameras (see Papers II and III). Therefore, the appropriate 
{effective wavelengths of the optical bandpasses} were used in the spectral index estimation. {Figure \ref{figTHREE} presents histograms of the broad-band spectral indices for ``core'' and ``power-law'' galaxies.}
As is clearly observed in Figure \ref{figTHREE}, the $\alpha_{ro}$ and $\alpha_{rx}$ {is} flatter (or more inverted) in the ``power-law'' galaxies compared to the ``core'' galaxies. This is {fully} consistent with the earlier finding that ``power-law'' galaxies show a deficit in the radio emission at a given X-ray (or optical) luminosity, as compared to {the} ``core'' galaxies.
{$\alpha_{ox}$ on the other hand, does not seem to differ much between the ``core'' and ``power-law'' galaxies (top right panel of Figure \ref{figTHREE}).} {We discuss the statistical significance of these results below.}

\begin{deluxetable}{ccccc}
\tabletypesize{\small}
\tablecaption{Two-point Spectral indices}
\tablewidth{0pt}
\tablehead{\colhead{Source} & \colhead{$\alpha_{ro}$} & \colhead{$\alpha_{ox}$} & \colhead{$\alpha_{rx}$}  & \colhead{SBP}\\
\colhead{(1)}&\colhead{(2)}&\colhead{(3)}&\colhead{(4)}&\colhead{(5)}}
\startdata
UGC\,0968  &  $>-$0.56 &  ...       &   $>-$0.26  & Core\\
UGC\,5617  &  $>-$0.67 &  $<-$0.04  &   $-$0.44   & PLaw\\
UGC\,5663  &  $-$0.70  &  0.55      &   $-$0.25   & PLaw\\
UGC\,5902  &  $>-$0.66 &  ...       &   $>-$0.33  & Core\\
UGC\,5959  &  $-$0.68  &  0.12      &   $-$0.39   & PLaw\\
UGC\,6153  &  $-$0.97  &  0.06      &   $-$0.55   & PLaw\\
UGC\,6297  &  $>-$0.52 &  ...       &   $>-$0.25  & Core\\
UGC\,6860  &  $>-$0.71 &  $>$0.07   &   $-$0.42   & Inter\\
UGC\,6946  &  $-$0.46  &  $-$0.22   &   $-$0.38   & PLaw\\
UGC\,6985  &  ...      &  ...       &   ...       & PLaw\\
UGC\,7005  &  $>-$0.59 &  $>$0.24   &   $-$0.30   & Inter\\
UGC\,7103  &  $-$1.06  &  0.59      &   $-$0.39   & PLaw\\
UGC\,7142  &  $-$0.61  &  0.04      &   $-$0.38   & PLaw\\
UGC\,7203  &  $-$0.44  &  $<$0.25   &   $>-$0.19  & Core\\
UGC\,7256  &  $-$0.63  &  0.08      &   $-$0.37   & PLaw\\
UGC\,7311  &  ...      &  ...       &   $-$0.37   & PLaw\\
UGC\,7360  &  $-$0.24  &  $-$0.11   &   $-$0.19   & Core\\
UGC\,7386  &  $-$0.17  &  $-$0.18   &   $-$0.17   & Core\\
UGC\,7494  &  $-$0.34  &  0.11      &   $-$0.16   & Core\\
UGC\,7575  &  ...      &  ...       &   $-$0.43   & Inter\\
UGC\,7614  &  $>-$0.86 &  ...       &   $>-$0.55  & PLaw\\
UGC\,7629  &  $-$0.14  &  0.09      &   $-$0.06   & Core\\
UGC\,7654  &  $-$0.30  &  0.19      &   $-$0.10   & Core\\
UGC\,7760  &  $-$0.18  &  $-$0.06   &   $-$0.14   & Core\\
UGC\,7797  &  $>-$0.40 &  ...       &   $>-$0.17  & Core\\
UGC\,7878  &  $-$0.41  &  $<$0.17   &   $>-$0.19  & Core\\
UGC\,7898  &  $>-$0.24 &  ...       &   $>-$0.13  & Core\\
UGC\,8355  &  $-$0.88  &  ...       &   ...       & PLaw\\
UGC\,8499  &  $>-$0.61 &  ...       &   ...       & Core\\
UGC\,8675  &  $-$0.96  &  0.04      &   $-$0.55   & Inter\\
UGC\,8745  &  ...      &  ...       &   $>-$0.18  & Core\\
UGC\,9655  &  ...      &  ...       &   ...       & Core\\
UGC\,9692  &  $>-$0.81 &  $>$0.49   &   $-$0.33   & PLaw\\
UGC\,9706  &  $-$0.33  &  0.26      &   $-$0.12   & Core\\
UGC\,9723  &  ...      &  ...       &   $>-$0.16  & Core\\
UGC\,10656 &  $>-$0.72 &  $>$0.17   &   $-$0.40   & PLaw\\
UGC\,12759 &  $-$0.98  &  0.58      &   $-$0.35   & PLaw\\
NGC\,1380  &  $>-$0.87 &  ...       &   ...       & PLaw\\
NGC\,1316  &  $>-$0.65 &  0.38      &   $-$0.23   & Core\\
NGC\,1399  &  $-$0.32  &  $>-$0.002 &   $>-$0.20  & Core\\
NGC\,3258  &  $-$0.51  &  ...       &   ...       & Core\\
NGC\,3268  &  $>-$0.24 &  ...       &   ...       & Core\\
NGC\,3557  &  ...      &  ...       &   $-$0.27   & Core\\
NGC\,4373  &  $-$0.33  &  ...       &   ...       & Core\\
NGC\,4696  &  $-$0.19  &  $-$0.16   &   $-$0.18   & Core\\
NGC\,5128  &  $-$0.54  &  $-$0.20   &   $-$0.39   & Core\\
NGC\,5419  &  $-$0.39  &  $-$0.11   &   $-$0.29   & Core\\
NGC\,6958  &  $-$0.69  &  ...       &   ...       & PLaw\\
IC\,1459   &  $-$0.22  &  $-$0.02   &   $-$0.15   & Core\\
IC\,4296   &  $-$0.21  &  $-$0.23   &   $-$0.22   & Core\\
IC\,4931   &  $-$0.54  &  ...       &   ...       & Core
\enddata
\tablecomments{\small Col. 1: Source name. Cols. 2, 3, 4: Radio-to-optical, optical-to-X-ray and radio-to-X-ray spectral indices of the galaxy nuclei, respectively. Col. 5: HST surface brightness profile classification as ``core'' (Core), ``power-law'' (PLaw) and ``intermediate'' (Inter).}
\label{tabalpha}
\end{deluxetable}

\subsubsection{Testing the statistical significance of the ``core$-$power-law'' division}
The two-sided Kolmogorov-Smirnov (KS) statistical test (implemented as KSTWO in IDL) indicates that the ratio of the radio to the X-ray flux density in the ``core'' and ``power-law'' galaxies from the previous sample ({\it i.e.,} excluding the 13 UGC galaxies {presented} here), was {\it not} statistically significantly different $-$ the KS test probability that the two data sets are drawn from the same distribution {was} 0.34. However, on the inclusion of the 13 new galaxies, the division between the ``core'' and ``power-law'' galaxies becomes statistically significant (KS test probability, $p$ = 0.049). This is also true for the radio emission alone: $p$ = 0.11 for the previous sample, but $p$ = 0.022 for the sample including the 13 new sources. (Note that all limits were assumed to be detections for the KS test.) 
{Our results, therefore,} strengthen the conclusions reached in \citet{Capetti05,BalmaverdeCapetti06,CapettiBalmaverde06}, {by placing them on a firmer statistical footing}. 

To take into account the upper limits in the radio, optical and X-ray flux {density} values, we examined the spectral index differences between the ``core'' and ``power-law'' galaxies with survival statistics or ASURV \citep{Isobe86,Lavalley92}, as implemented in the IRAF package. We used the ``twosampt" task in ASURV, which computes several nonparametric two-sample tests for comparing two or more censored data sets. These tests revealed that while $\alpha_{ro}$ and $\alpha_{rx}$ were significantly different between the ``core'' and ``power-law'' galaxies, $\alpha_{ox}$ was not. For instance, the Gehan's Generalized Wilcoxon test estimates the probability for the two sets of censored data for $\alpha_{ro}$ to belong to the same population is $<10^{-5}$ (test statistic = 4.252), and for $\alpha_{rx}$ is $<10^{-5}$ (test statistic = 4.630). The Gehan's Generalized Wilcoxon test indicates a $p=0.249$ (test statistic = 1.152) for the $\alpha_{ox}$ data. {We therefore conclude that the ``core'' and ``power-law'' galaxies differ primarily in their radio outflow properties.} 

\section{Discussion}
In Papers I, II and III, it was found that ``core'' and ``power-law'' galaxies occupied different regions in the radio-X-ray plane. These results gain a much firmer statistical footing with the inclusion of 13 new galaxies presented in this paper. The galaxies which are classified as having ``intermediate'' surface brightness cusp slopes, lie in the region occupied by the ``power-law'' galaxies; although more of them are needed to confirm this trend. 

The new VLA data presented here has ten times better spatial resolution than previous radio observations of the sample. Even with these high resolution radio observations, weak but compact emission is detected in the centers of all but one galaxy. This supports the picture of an AGN in these low luminosity galaxies, although multifrequency radio data are required to derive radio spectral indices and truly ascertain their AGN nature. 

Our multiwavelength study points out that ``core/power-law'' (or equivalently, radio-loud/radio-quiet) galaxies primarily differ in their radio {outflow} properties. Relatively more powerful radio outflows can be launched in ``core'' galaxies compared to ``power-law'' galaxies. 
{There appear to be no statistically significant differences between the black hole masses of ``core'' and ``power-law'' galaxies \citep[Papers II, III,][]{Ho02};} although the ``core'' galaxies {do} possess black holes with masses that are typically $\gtrsim10^{8}$\,M$_{\sun}$ (see Papers II, III). The {suggestion} of differences in the spins of the central black holes has been put forth to explain the ``radio-loud/radio-quiet'' dichotomy \citep[e.g.,][]{Sikora07} $-$ central black holes in giant elliptical galaxies are suggested to have much larger spins on average, than black holes in spiral or disk galaxies {that host radio-quiet AGN}. However, this dichotomy breaks down at high accretion rates, because the dominant fraction of quasars in giant elliptical galaxies are radio-quiet. Jet intermittency is therefore suggested in sources with high accretion rates. \citet{Chiaberge11} have proposed that the spin of the black hole plays an instrumental role in making a source radio-loud {\it only} if its black hole mass is $\gtrsim10^{8}$\,M$_{\sun}$. This lower limit is indeed consistent with the typical black hole masses of the ``core'' galaxies in our sample (Papers II, III).

{Recently}, \citet{Richings11} {have} claimed that the correspondence between the ``radio-loud/radio-quiet'' and the ``core/power-law'' classification {is} not as clean as previously claimed. Richings et al. have considered nearby galaxies from the Palomar survey \citep{Ho95}. This is justified because imposing a radio selection could bias the sample (e.g. by excluding radio-quiet AGN in ``core'' galaxies). However, they include all galaxies with Hubble type $T\le3$, and therefore also include {\it Sa} and {\it Sb} type spiral galaxies (in total about 1/2 of their sample), and not just early-type galaxies, like we do. In addition, they measure radio loudness using the ratio between the radio and [OIII] luminosity. However, in low luminosity AGN, [OIII] is not a good measure of the AGN power because it is contaminated by other (galactic) sources \citep[e.g.,][]{Capetti11}. About half of their sample is below the threshold power of a few $10^{38}$ erg~sec$^{-1}$, that could be trusted as being  AGN. Indeed, when they limit their analysis to the galaxies with optical nuclei, they recover the clean separation that we observe. 

The ``core/power-law'' galaxy differences possibly indicate different formation histories and evolution in them. For example, a ``core'' galaxy could be the result of (at least) one major merger, and the core formation could be related to the dynamical effects of the binary black holes \citep[e.g.,][]{Milosavljevic02}. Conversely, ``power-law'' galaxies partly preserve their original disky appearance, suggestive of a series of minor mergers \citep{Faber97}. 
This is consistent with the picture outlined, for example, by \citet{Merritt06,Kormendy09,Chiaberge11}: ``core'' galaxies most likely originate in major gas-deficient ``dry'' mergers, while ``power-law'' galaxies may originate in gas-rich ``wet'' mergers. The binary black holes formed in the former, would eject stars away from the central regions and produce the observed starlight deficit, while the starburst triggered in the latter, would create the additional starlight in the centers and ``fill'' the cores. Our results are consistent with the suggestion that the same processes also determine the characteristic of the active nucleus. In the merger process, the SMBH associated with each galaxy rapidly sinks toward the center of the forming object. Thus, the resulting nuclear configuration after the merger (described by the total mass, mass ratio or separation of the SMBHs) is directly related to the evolution of the host galaxy. \citet{WilsonColbert95} have suggested that a radio-loud AGN could form only after the coalescence of two SMBH of similar (large) mass, forming a highly spinning nuclear object, from which the energy necessary to launch a relativistic jet could be extracted. Minor mergers in radio-quiet AGN could result in the formation of short-lived accretion disks and radio outflows in random directions with respect to the symmetry axes of the galaxies \citep[e.g., in Seyfert galaxies,][]{Kharb06,Kharb10b}, which could gradually spin down the supermassive black holes. 

\section{Summary \& Conclusions}
\begin{enumerate}
\item We have observed 13 low luminosity UGC galaxies with {\it Chandra}-ACIS and the VLA A-array configuration at 4.9 GHz. This completes the multiwavelength study of a sample of 51 nearby early-type galaxies described in Papers I, II, and III. Nuclear X-ray emission is detected in only eight galaxies. Compact radio emission is detected in all sources except UGC\,6985. The compact radio emission in these galaxies is strongly supportive of their AGN nature. 
\item The 13 UGC galaxies lie in distinctly different regions of the radio-X-ray plane depending on their optical surface brightness profiles, just like the rest of the sample galaxies. While the ``core'' galaxies lie along the extrapolation of the established FRI radio galaxy correlation, the ``power-law'' galaxies lie above the FRI correlation. The four ``intermediate'' galaxies appear to lie in the region occupied by  the ``power-law'' galaxies. The Kolmogorov-Smirnov test on the radio-to-X-ray flux density ratio indicates that the addition of the 13 new sources puts the previous results on a much firmer statistical footing $-$ the KS test probability $p$ of the two classes being statistically similar changes from 0.34 to 0.049, when the 13 new galaxies are included.
\item Broad-band (radio-optical-X-ray) spectral indices are presented for all 51 sources. Survival statistics (ASURV) indicates that  
{the ``core'' and ``power-law'' galaxies have significantly different radio-to-optical and radio-to-X-ray spectral indices but similar optical-to-X-ray spectral indices. The Gehan's Generalized Wilcoxon test yields a probability $p<10^{-5}$ that the radio-to-optical and radio-to-X-ray spectral indices are drawn from the same populations, whereas $p= 0.25$ for the 
optical-to-X-ray spectral index. } 
This supports the idea that relatively powerful radio outflows are launched more easily in ``core'' galaxies compared to ``power-law'' galaxies.
\item {Overall,} these results point to different evolution and formation histories in ``core'' and ``power-law'' galaxies: major mergers are likely to have created ``core'' galaxies, while minor mergers were instrumental in the creation of ``power-law'' galaxies.
\end{enumerate}

\acknowledgments We thank the referee for making useful suggestions that have improved this paper. We thank Henrique Schmitt for providing us with the radio fluxes of two sources. This work was supported by NASA grant GO6-7089X. The National Radio Astronomy Observatory is a facility of the National Science Foundation operated under cooperative agreement by Associated Universities, Inc.  This research has made use of the NASA/IPAC Extragalactic Database (NED) which is operated by the Jet Propulsion Laboratory, California Institute of Technology, under contract with the National Aeronautics and Space Administration.

{\it Facilities:} \facility{CXO (ACIS)}, \facility{VLA}.

\appendix
\section{Notes on individual galaxies}
{\bf UGC\,0968 (NGC\,0524):} This massive S0 galaxy has a relatively rich globular cluster system and dominates a small galaxy group \citep{Larsen01}. Its HST image reveals the presence of concentric dust rings \citep{Ravindranath01}. {The} {\it Chandra} image does not indicate the presence of a compact X-ray core in this galaxy. We did not observe this source with the VLA.

{\bf UGC\,5959 (NGC\,3414):} This ``peculiar'' galaxy has a dominant central bulge and a rudimentary disk of very low surface brightness. A sharply tipped S-shaped feature, together with a large twist, is interpreted as evidence for a spiral pattern in the disk by \citet{Michard94}. It forms the close pair with NGC\,3418. {The} {\it Chandra} image reveals the presence of an X-ray nucleus in this galaxy. We did not observe this source with the VLA.

{\bf UGC\,6860 (NGC\,3945):} This S0 galaxy has been classified as being double-barred with a large inner disk dominating the region between the two bars \citep{Erwin99}. The {\it Chandra} and VLA images reveal the presence of an X-ray/radio nucleus in this galaxy.

{\bf UGC\,6946 (NGC\,3998):} This galaxy has been classified as an AGN-dominated LINER \citep{Alonso00}. The nucleus contains a compact flat-spectrum radio source \citep{Hummel80}. Judging from the large difference in flux density between our VLA measurements and those of \citet{wrobel91a}, strong variability is suspected in the nucleus of this galaxy. \citet{Ford86} have presented a narrow-band H{$\alpha$} image in which a 90$\arcsec$ long S-shaped structure is seen positioned intermediate between the major- and minor-axis. The {\it Chandra} and VLA images of this source reveal a bright nucleus.

{\bf UGC\,6985 (NGC\,4026):} This galaxy is an edge-on S0 displaying a dominant disk with an inner ring and an outer lens \citep{Michard94}. It is not clear if there is a compact X-ray nucleus in this galaxy. No VLA radio emission is detected in this source. This is consistent with the lack of an AGN in this galaxy.

{\bf UGC\,7005 (NGC\,4036):} This galaxy is a disk dominated S0 with an irregular dust lane threading the disk \citep{Michard94}. It forms a non-interacting pair with NGC\,4041 at 15$\arcmin$. This galaxy has been classified as harboring a LINER nucleus \citep{Fisher97}. The {\it Chandra} image reveals an X-ray nucleus along with diffuse emission. The VLA image reveals jet-like extended radio emission at a position angle similar to the X-ray emission extension in this source, although on a much smaller spatial scale. The diffuse X-ray emission in fact extends roughly along the position angle of the galactic disk in this source. 

{\bf UGC\,7311 (NGC\,4233):} This galaxy is classified as an SB0. The elongation of the central, nearly circular dominant bulge, at an angle of about 70$\degr$ to the major axis of the flattened envelope, accounts for this classification. 
The presence of a compact nucleus is indicated both in {the} {\it Chandra} and VLA images.

{\bf UGC\,7797 (NGC\,4589):} This galaxy is suggested to be a merger remnant on the basis of its complex gas and stellar kinematics \citep{Ravindranath01}. Dust filaments traverse the center of the optical image. This galaxy forms a non-interacting pair with NGC\,4572 at 7$\farcm$5. This is classified as a LINER based on the [O II] to [O III] ratio \citep{Larkin98}. {The} {\it Chandra} image does not indicate the presence of a clear nucleus in this galaxy, but the VLA image does reveal radio emission.

{\bf UGC\,8499 (NGC\,5198):} This galaxy is classified as an E/S0. The {\it Chandra} image does not indicate the presence of a clear nucleus in this galaxy, but the VLA image does reveal radio emission. \citet{Krajnovic02} point out that there is no nuclear dust structure in this galaxy.

{\bf UGC\,8745 (NGC\,5322):} The {HST} image of this galaxy displays a completely obscuring dust lane, which is probably a nearly edge-on disk perpendicular to the broad radio jet and cuts the nucleus in two and hides the very center of the galaxy \citep{Carollo97}. It has been classified as a LINER or a Seyfert 2 \citep{Ho97}. {The} {\it Chandra} image does not indicate the presence of a clear nucleus in this galaxy, but diffuse X-ray emission is present. The VLA image reveals a clear bipolar jet-like radio emission in this galaxy \citep[also see][]{Feretti84}. The radio emission extension is nearly perpendicular to the X-ray emission extension in this galaxy.

{\bf UGC\,9692 (NGC\,5838):} There is evidence of a spiral pattern in the outer regions of the disk of this strongly inclined S0 galaxy. Its HST image reveals a thick nuclear dust ring \citep{Ravindranath01}. The {\it Chandra} image may be indicating the presence of a nucleus. A lot of diffuse X-ray emission is also present. The VLA image reveals jet-like extended radio emission at a position angle similar to the X-ray emission extension in this source.

{\bf UGC\,10656 (NGC\,6278):} This galaxy is classified as an S0, and is a member of a galaxy pair \citep{Peterson79}. {The} {\it Chandra} and VLA images of this source reveal nuclear emission.

{\bf UGC\,12759 (NGC\,7743):} This galaxy is classified as SBa or S0. Its HST image reveals the presence of several dust lanes with large pitch angle that imply the same sense of rotation, although these dust lanes do not appear to extend more than approximately 500 pc from the nucleus \citep{Martini03}. It has a LINER/Seyfert 2 type nucleus \citep{Ho97}. {The} {\it Chandra} image indicates the presence of an X-ray nucleus and some diffuse emission in this galaxy. We did not observe this source with the VLA.

\bibliographystyle{apj}
\bibliography{ms}

\end{document}